\documentclass[final,5p,times,twocolumn]{elsarticle}

\usepackage{lineno,hyperref}
\usepackage{hhline}
\usepackage{array}
\usepackage{multirow}
\usepackage{booktabs}
\usepackage{tabularx,hhline}
\usepackage{float}
\usepackage{amsmath}
\usepackage{gensymb}
\usepackage{color,soul} %to higlight text
\usepackage{mathtools, cuted}
\usepackage{lipsum, color}
\usepackage{lipsum}
\usepackage{graphicx}
\usepackage{subcaption}

\newcommand{\sipm}{SiPM}

\hypersetup{
    colorlinks=true,       % false: boxed links; true: colored links
    linkcolor=blue,          % color of internal links (change box color with linkbordercolor)
    citecolor=green,        % color of links to bibliography
    filecolor=magenta,      % color of file links
    urlcolor=cyan           % color of external links
}
\graphicspath{{images/}{./}}
\modulolinenumbers[5]

\journal{Journal of \LaTeX\ Templates}

\begin{document}

\begin{frontmatter}

\title{Characterisation of a large area silicon photomultiplier}

%% Group authors per affiliation:
\author[UnigeAddress]{A. Nagai \corref{mycorrespondingauthor}}
\cortext[mycorrespondingauthor]{Corresponding author}
\ead{Andrii.Nagai@unige.ch}
\author[UnigeAddress]{C. Alispach}
\author[UnigeAddress]{A. Barbano}

\author[CERN]{V. Coco}
%\ead{victor.coco@cern.ch}

\author[UnigeAddress]{D. della Volpe}
%\ead{Domenico.Della.Volpe@cern.ch}
\author[UnigeAddress]{M. Heller}
%\ead{Matthieu.Heller@cern.ch}

\author[UnigeAddress]{T. Montaruli}
%\ead{Teresa.Montaruli@cern.ch}
\author[UnigeAddress]{S. Njoh}

\author[UnigeAddress]{I. Troyano-Pujadas}

\author[UnigeAddress]{Y. Renier}

\address[UnigeAddress]{D\'epartment de physique nucl\'eaire et corpusculaire, Universit\'e de Gen\`eve, 24 Quai E. Ansermet, CH-1211, Switzerland}
\address[CERN]{European Organization for Nuclear Research CERN, 1, Esplanade des particules, CH-1211 Genève 23, Switzerland}

%% or include affiliations in footnotes:

\begin{abstract}
This work illustrates and compares some  methods to measure the most relevant parameters of silicon photo-multipliers (\sipm{}s), such as photon detection efficiency as a function of over-voltage and wavelength, dark count rate, optical cross-talk, afterpulse probability. 
For the measurement of the breakdown voltage, $V_{BD}$, several methods using the current-voltage $IV$ curve are compared, such as the ``IV Model", the ``relative logarithmic derivative", the ``inverse logarithmic  derivative", the ``second logarithmic derivative", and the ``third derivative" models.
We also show how some of these characteristics can be quite well described by few parameters and allow, for example, to build a function of the wavelength and over-voltage describing the photodetection efficiency. This is fundamental to determine the  working point of SiPMs in applications where external factors can affect it.

These methods are applied to the large area monolithic hexagonal SiPM S10943-2832(X), developed in collaboration with Hamamatsu and  adopted for a camera for a gamma-ray telescope, called the SST-1M. 
We describe the measurements of the performance at room temperature of this device. The methods used here can be applied to any other device and the physics background discussed here are quite general and valid for a large phase-space of the parameters.

%Also, the drop of photodetection efficiency ($PDE$) with night sky background ($NSB$) is investigated.

\end{abstract}

\begin{keyword}
\sipm{}, MPPC, $PDE$, $V_{BD}$, cross-talk, dark count rate, afterpulses, triggering probability
\MSC[2010] 00-01\sep  99-00
\end{keyword}

\end{frontmatter}

%\linenumbers

\section{Introduction}

In the last few years the interest in solid state photodetectors has grown significantly.
In particular, SiPMs~\footnote{Hamamatsu adopted the name Multi-Pixel Photon Counters or MPPCs} have replaced traditional photo-multiplier tubes (PMTs) in many applications. As a matter of fact, they are very compact, robust, lightweight, insensitive to magnetic fields and work at temperatures that span a wide range from cryogenics to beyond room temperature.
Their operating parameters are stable across devices of the same type thanks to the high level of uniformity achieved by the solid state technology production technique.
Also the absence of aging caused by the integrated light over time, makes them particularly tailored for ground-based astrophysics~\citep{DuneAging}, where they can be operated even in the presence of high background light level, thus increasing the duty cycle and then the physics reach of experiments~\citep{FACT}.

The University of Geneva and a Consortium of Polish and Czech Institutions have proposed and built a single mirror small size telescope (SST-1M) for the Cherenkov Telescope Array (\href{https://www.cta-observatory.org}{CTA}), equipped with a SiPM-based camera. 
To achieve the desired performance with the chosen optics, a mirror of 4~m diameter, the camera of the SST-1M is composed by 1296 pixels, each of an angular opening of about $0.24^\circ$. This translates into a pixel linear size of about 2.32 cm (more details on camera and its design and performances can be found in ref.~\citep{CameraPaperHeller2017}).
 
In order to have a spatial uniform response of the camera, the pixels should have a circular shape to ensure equal distance between pixel centres in every direction. 
The hexagon is the best possible shape to achieve this uniformity with minimum dead space. 
The pixel size to achieve the required angular resolution is achieved through a large SiPM coupled with light funnel. 
%Given the large pixel size, it is mandatory to use a light funnel to reduce the size of the sensor, both for cost reason and performance.
A light funnel, approaching the ideal Winston cone geometry, was designed by the 
University of Geneva group to be coupled to \sipm{}s and achieve the desired pixel size.
The light funnel has hexagonal shape and has a compression factor of about six~\cite{Aguilar:WinstonCones_2014}. Its internal surface is coated in order to maximise reflection of UV Cherenkov light produced by the cosmic rays when traversing the atmosphere, and also to have a good reflectivity for light with a direction almost parallel to cone surface.

The Winston cone geometry, on the other side, imposes to have the same shape at entrance and exit side and then an hexagonal sensor was needed.
This was developed by the University of Geneva group in cooperation with the Hamamatsu company (Hamamatsu S10943-2832(X)). 
The main characteristics of the sensor are detailed in Tab.~\ref{tab:HexSiPM}. The sensor area is around 93.6 mm$^2$ with a linear dimension of 10.4 mm flat-to-flat. It ranks among the world's largest monolithic sensor.
The large area can be a limiting factor in many application. 
As matter of fact, the capacitance and the dark-count rate ($DCR$) are proportional to \sipm{} area. Hence, larger devices tend to have longer output signals and be more noisy.  
However, as shown by the SST-1M camera~\cite{CameraPaperHeller2017}, with the proper electronics, such a large device can achieve the desired performances in specific applications.

This work reports on the characterization studies done to validate the design and verify the performances of the SST-1M new sensor type. 
\begin{table}[hbt]
  \centering
  \renewcommand{\arraystretch}{1.4}
\begin{tabular}{||r|c||}
\hhline{|t:==:t|}
Nr. of channels & 4 \\
Cell size & 50 $\times$ 50 $\mu$m $^{2}$ \\
Nr of cells (per channel)         & 9210    \\
Fill Factor	       & 61.5\%  \\
$DCR$ (@$V_{op}$ per channel) 	& 2.8-5.6 MHz\\
$C_{\mu cell}$ (@ $V_{op}$ per channel) & 	85  fF\\
Cross-talk (@$V_{op}$  per channel)	&  10\%\\
$V_{BD}$ Temp. Coeff.  & 54 mV/C$^\circ$\\
Gain (@$V_{op}$  per channel) &	$1.49 \times10^6$\\
\hhline{|t:==:t|}
  \end{tabular}
\caption{S10943-2832(X) \sipm{} main characteristics provided by the producer at T = 25 $^{\circ}$C. $V_{op} = V_{BD} + 2.8$ V.}
\label{tab:HexSiPM}
\end{table}

\section{The Hamamatsu S10943-2832(X) \sipm{}}
\begin{figure}[ht]
\begin{center}\includegraphics[%
  width=0.8\linewidth,
  keepaspectratio]{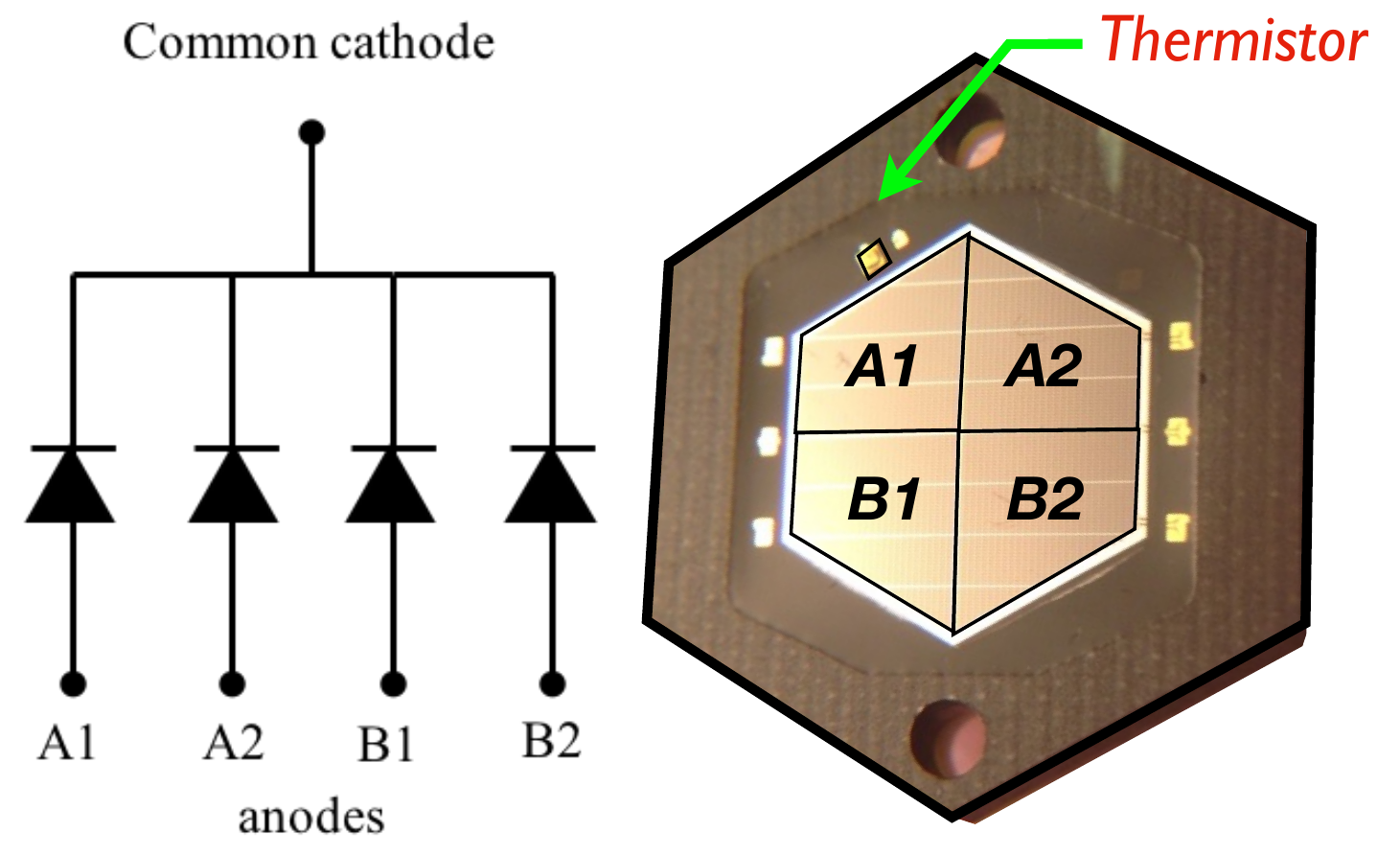}\end{center}
\caption{Picture of the Hamamatsu S10943-2832(X) \sipm{} (right) and its electric equivalent model (left). On the sensor package also a NTC temperature probe is present. 
This is used to monitor the temperature variation  affecting  parameters as DCR or V$_{breakdown}$, for which a real-time correction can be applied to keep the working point
stable~\cite{SST1Melectronics}. 
}
\label{Fig:HexSensor}
\end{figure}

The \sipm{} S10943-2832(X), shown in Fig.~\ref{Fig:HexSensor}, has been designed in collaboration with Hamamatsu and is based on the so called LCT2 (Low cross-talk) technology available when the camera design was done. Hamamatsu has further improved this technology (LCT5 or LVR) and offers now better performance. 
It is worth to mention that the hexagonal shape can be obtained using any Hamamatsu $\mu$cell standard technology and size, through a dedicated photo-mask. 
Currently, we evaluate that the slightly higher DCR of LCT2 does not impact significantly performances of camera if appropriately calibrated. 
Actually, dark counts are useful for
in-situ calibrations and then a further reduction of this rate increases the time needed to accumulate the statistics needed for precise calibrations.

\begin{figure}[ht]
\begin{center}\includegraphics[%
  width=0.8\linewidth,
  keepaspectratio]{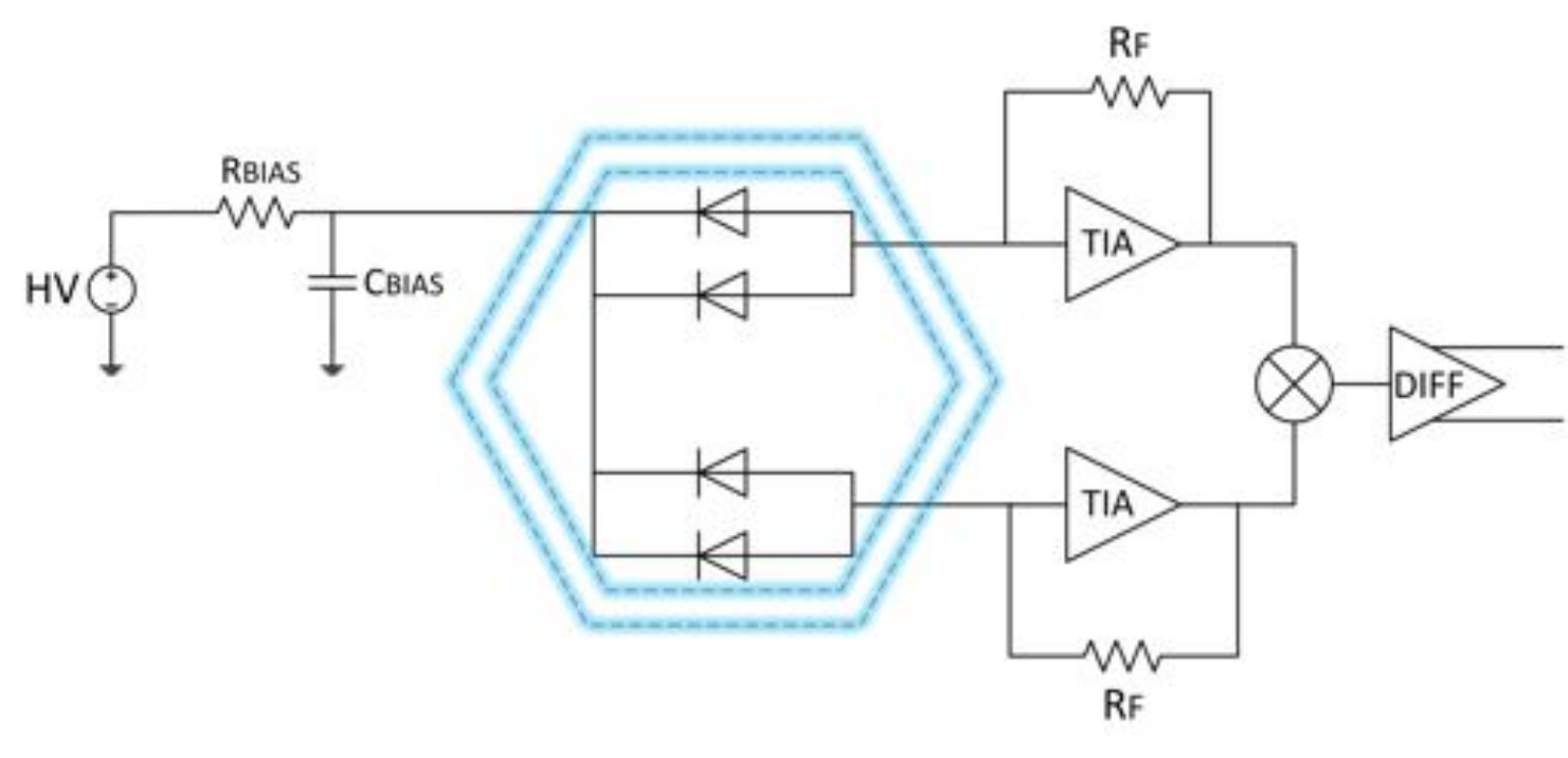}\end{center}
\caption{\sipm{} polarisation scheme and pre-amplifier topology used to sum up its four channels.}
\label{Fig:PreAmpTopology}
\end{figure}
The sensor capacitance is directly related to its active area and this has an impact on the signal recharge time. 
In this case,
signals would have typical duration of about hundred ns, a too  long time for the desired bandwidth of 250 MHz. 
This frequency has been chosen taking into account the typical time duration of atmospheric showers induced by gamma-rays and cosmic rays.
To reduce the effect of the capacitance, the sensor has four independent anodes and a common cathode as shown in Fig.~\ref{Fig:HexSensor}.
This configuration allows to readout the 4 channels independently but there is a single bias for the whole sensor.
Nonetheless, in order to achieve the desired bandwidth, a shaping of the signal is needed.

To address this features, we developed in house the pre-amplification chain based on off-the-shelf components. 
The solution adopted~\citep{SST1Melectronics} is a trans-impedance amplifier topology with low noise amplifiers (\href{http://www.ti.com/lit/ds/symlink/opa846.pdf}{OPA846}) as it can achieve the required events rate with the best signal-to-noise ratio and gain/bandwidth ratio. 
As shown in Fig.~\ref{Fig:PreAmpTopology}, the four channels are summed by two in order to reduce the equivalent capacitance and pulse length. The summed signals are further summed up in a differential amplifier, which feeds the output signal into the digital readout system. 

Another important characteristic of the camera architecture is the fact that the front-end and the digital readout are DC coupled. This is important for gamma-ray astronomy, where Moon light and human-induced light and their reflections are a relevant background. As a matter of fact, the Night Sky Background (NSB) contributes to the determination of the real working point of the device, relevant to correctly extract the number of photons from the signal.

\section{Static characterisation}

All the laboratory measurements (i.e. static, dynamic and optical) are performed at room temperature T = 25 $^{\circ}$C at the premises of IdeaSquare\footnote{\url{http://ideasquare.web.cern.ch}} at CERN, where an experimental setup has been developed.
The static characterization (i.e. reverse and forward current-voltage (IV) curves), is performed using a Keithley 2400~\cite{Keithley2400} pico-ammeter for bias supply and current measurements.

\subsection{Forward IV characterization}
\label{sec:ForwardIV}

\begin{figure}[ht]
\begin{center}\includegraphics[%
  width=8.2cm,
  keepaspectratio]{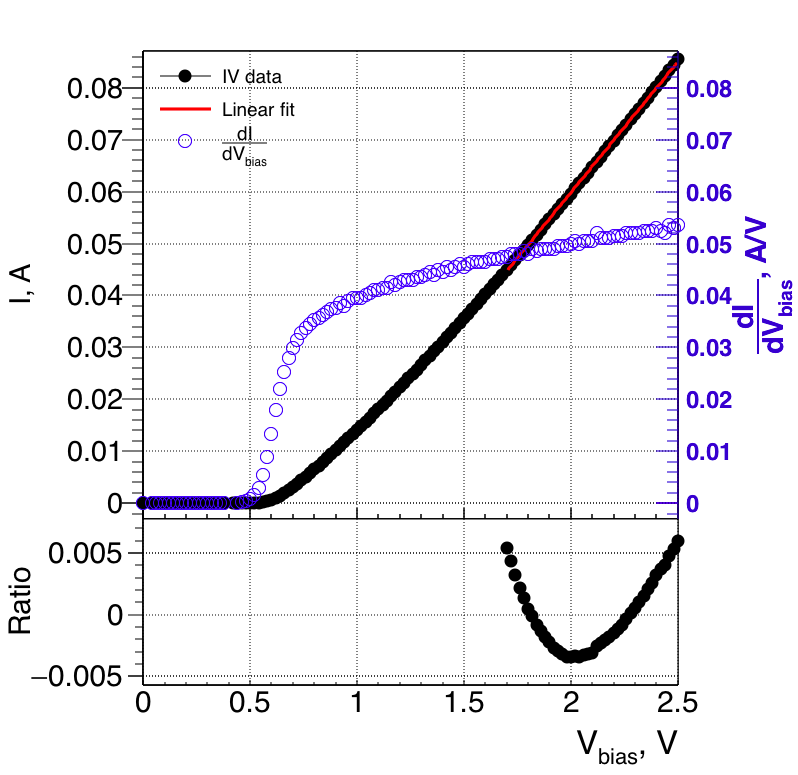}\end{center}
\caption{The forward IV characteristic and its derivative of the Hamamatsu S10943-2832(X) \sipm{}. The linear fit (red line) is superimposed to data points. In the bottom pane is shown the $Ratio = \left( I_{data} - I_{fit} \right)/ I_{data}$.}
\label{Fig:IVForward}
\end{figure}

The forward IV characteristic curve of the \sipm{}, shown in Fig.~\ref{Fig:IVForward}, exhibits a very small increase of the current when the polarization voltage, $V_{bias}$, is below the threshold value and a linear rapid current increase with $V_{bias}$ above this threshold.
A physical interpretation of this behaviour can be attempted starting from the ideal Shockley law ~\citep{Shockley1949}, which expresses the forward current flowing $I^{d}$ through a \textit{p-n} diode as:

\begin{equation}
 \label{eq1}
I^{d} = I_{s}^{d} \left[ exp \left(  \frac{V_{j}}{ \eta V_{T}} \right ) - 1 \right]  ,
\end{equation}
where $I_{s}^{d}$ is the diode reverse bias saturation current, $V_{j}$ is the voltage across the junction, $V_{T}$ is the thermal voltage and $\eta$ is the ideality factor. The voltage $V_{j}$ is the difference between the applied voltage $V_{bias}$ and the voltage drop across the neutral region and the ohmic contacts on the two sides of the junction:
\begin{equation}
 \label{eq2}
V_{j} = V_{bias} - I^{d} \cdot R_{s}
\end{equation}
where usually $R_s\simeq$100 $\Omega$.

Replacing $V_{j}$ by $V_{bias}$ in Eq.~\ref{eq1}, we obtain:
\begin{equation}
 \label{eq3}
V_{bias} =  \eta V_{T}  \left[ \ln \left( \frac{ I^{d} }{ I_{s}^{d} } + 1  \right)  \right] + I^{d} \cdot R_{s} \, .
\end{equation}

The \sipm{} is an array of $N_{\mu cell}$ micro-cells ($\mu$cells), which are SPADs (single photon avalanche diode). Each $\mu$cell can be represented by a diode connected in series with a quenching resistor $R_{q}$\footnote{The $\mu$cell works in Geiger-Avalanche mode meaning that when a photon is absorbed, an electron-hole pair is created and the high electric field in the junction starts charge multiplication, which produces an avalanche. If the field is not reduced, the charge avalanche is stationary and leads to thermal destruction of the device. By adding a resistor in series to the $\mu$cell, a voltage drop is produced by the current induced by the charge avalanche when flowing into the resistor. This drop reduces the field across the device thus quenching the avalanche. For this reason the \sipm{} are also referred to as an array of G-APDs - Geiger-Avalanche Photo-Diodes.}.
Eq.~\ref{eq3} applies to each single $\mu$cell but requires the addition of the voltage drop caused by the presence of a quenching resistance $R_{q}$.
%:
%
%\begin{equation}
% \label{eq4}
%V_{bias} =  \eta V_{T}  \left[ exp \left(  \frac{ I }{ I_{s} } \right ) + 1 \right ] + IR_{s} + IR_{q}
%\end{equation}
%
%or
Then for a full SiPM device with $N_{\mu cell}$ connected in parallel, Eq.~\ref{eq3} becomes:

\begin{equation}
 \label{eq5}
V_{bias} =  \eta V_{T}  \left[ \ln \left(  \frac{ I }{ I_{s} }   + 1 \right) \right] + I \frac{(R_{s} + R_{q})}{ N_{\mu cell} } \, .
\end{equation}
where  $I_{s}$ is the \sipm{} reverse bias saturation total current and $I$ is the forward total current flowing through it.

The last term of Eq.~\ref{eq5}
%$ I \frac{R_{s}}{ N \mu cell } + I \frac{R_{q}}{ N \mu cell } $ term
becomes dominant when the current is high \mbox{($I/N_{\mu cell} >$ 5 $\mu$A)}. In this regime, \mbox{$R_{s}$ + $R_{q}$} can be extracted from a linear fit of the forward IV characteristic curve in Fig.~\ref{Fig:IVForward}:

\begin{equation}
 \label{eq6}
  R_{q} +  R_{s} = \frac{N_{\mu cell}}{ b} \underset{\vert R_{q} >> R_{s} \vert} \simeq  R_{q} \, .
\end{equation}
where $b$ is the slope parameter extracted by the linear fit (red line in Fig.~\ref{Fig:IVForward}). Also, $b$ can be calculated as $b = \frac{dI}{dV_{bias}}$.
For this \sipm{}, the fit gives  
$R_{q} = 182.9 \pm 0.3$ (stat.) $\pm 31$ (sys.)  k$\Omega$. 
The systematic uncertainty comes from the fact that $I$ does not increase linearly with $V_{bias}$. This can be seen from the bottom part of Fig.~\ref{Fig:IVForward}, showing $Ratio = \left( I_{data} - I_{fit} \right)/ I_{data}$ and $\frac{dI}{dV_{bias}}$. It is  calculated as:
\begin{equation}
 \label{eq7}
  \sigma_{sys.}^{R_{q}} = 0.5 \cdot \left( \frac{N_{\mu cell}}{ b_{1.6V}} - \frac{N_{\mu cell}}{ b_{2.5V} } \right) .
\end{equation}
where $b_{1.6V}$ and $b_{2.5V}$ are two slopes calculated at $V_{bias}$ of 1.6 V and 2.5 V respectively.
%The systematic uncertainty comes from the instrument quoted accuracy~\cite{Keithley2400} but does not account for the uncertainty on the series resistor $R_{s}$ which is of the order of 5-10\%.

\subsection{Reverse IV characterisation}
\label{sec:ReverseIV}

The current flowing in the \sipm{}, when not illuminated, depends on the available free carriers.
The  Shockley-Read-Hall (SRH) ~\cite{ShockleyReadPhysRev.87.1952,Hall1879} effect is the dominant one in semiconductors and it is also the main contribution to the bulk dark current.
It describes the generation and recombination of electron-hole pairs due to the trapping effect of impurities in the lattice (for this also called trap-assisted recombination), as well as band-to-band tunneling effects. 
In addition the carriers generation rate can be enhanced by reduction of activation energy due to the Poole-Frenkel effect \cite{PooleFrankel1938}.

\begin{figure}[ht]
\begin{center}\includegraphics[%
  width=8.2cm,
  keepaspectratio]{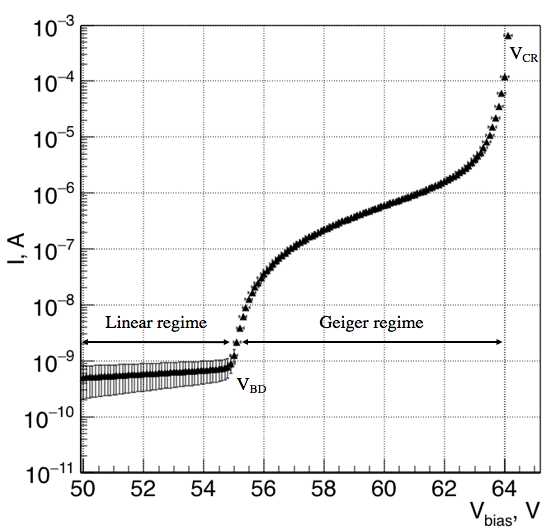}\end{center}
\caption{The reverse IV characteristic data points of the Hamamatsu S10943-2832(X) \sipm{}. Two different regions can be distinguished: $pre-breakdown$ and $post-breakdown$.}
\label{Fig:IVreverse}
\end{figure}

In the  reverse IV characteristic curve of the \sipm{}, shown in Fig.~\ref{Fig:IVreverse}, two zones are identified, corresponding to different regimes: 
\begin{itemize}
\item[(1)] the ``Linear'' regime (pre-breakdown), corresponding to $V_{bias}$ below the breakdown voltage ($V_{BD}$), where the current increases slowly with $V_{bias}$. This dark current is due to the surface current and the bulk dark current due to the free carriers.
\item[(2)] the ``Geiger'' regime (post-breakdown), corresponding to $V_{bias}$ above $V_{BD}$ where the current increases much faster with $V_{bias}$. This trend is due to the Geiger avalanche created by the free carriers generated by ionization. Primary free carriers, which trigger an avalanche, are usually created, due to the SRH thermal generation enhanced by Poole-Frenkel effect and tunnelling, but also to other associated effects as afterpulsing, prompt cross-talk and delayed cross-talk.
\end{itemize}

The breakdown voltage $V_{BD}$ of a \sipm{} device represents the voltage above which the electrical field inside the depleted region of a $\mu$cell is high enough that any free carrier (created by an absorbed photon or by a thermally generated carrier) can trigger an avalanche.
It marks, then, the transition between the two regimes and that is why it represent one of the most important parameter to determine.

The reverse IV measurements is commonly used for fast calculation of $V_{BD}$ using different methods such as the ``relative logarithmic derivative''~\cite{HamamatsuBook}, the ``inverse logarithmic derivative''~\cite{InvRelDerivativeMethod}, the ``second logarithmic derivative''~\cite{2ndDerivativeMethod}, the ``third derivative''~\cite{3rdDerivativeMethod} and ``IV Model'' methods~\cite{IVModeleMethod1,IVModeleMethod2}.
%A brief review of each model as well as the comparison of results obtained with the different methods are presented below.

%In the ``relative logarithmic derivative" and `inverse logarithmic derivative" methods, it is assumed that the dark count rate and cross-talk probability are linearly depending on the bias voltage, so that the \sipm{} gain increases linearly with the over-voltage $\Delta V=V_{bias}-V_{BD}$.

\begin{figure}[tb]
\begin{center}\includegraphics[%
  width=8.2cm,
  keepaspectratio]{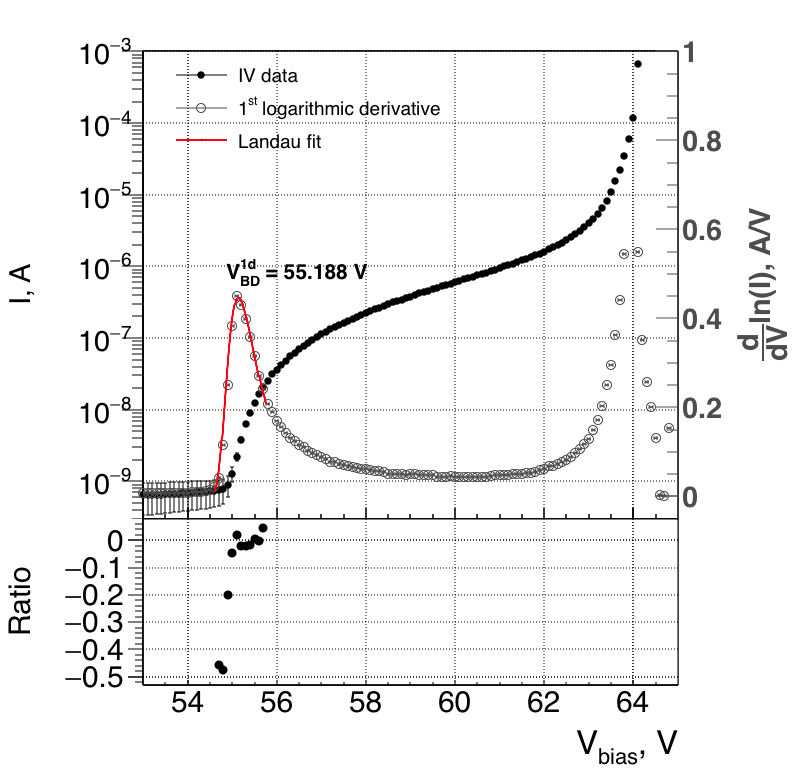}\end{center}
\caption{The reverse IV characteristic data points and its logarithmic derivative. A $V_{BD}^{1d} = 55.188$~V is extracted from the fit of the data (red line) with a Landau function. Also, the $Ratio = \left(  \frac{\mathrm d}{\mathrm dV} \ln(I) - I_{fit} \right) / \frac{\mathrm d}{\mathrm dV} \ln(I) $ is shown at the bottom of the figure.}
\label{Fig:FirstDerivative}
\end{figure}

In the ``relative logarithmic derivative" the breakdown voltage can be calculated~\cite{RelDerivativeMethod} as the voltage where $\frac{\mathrm d}{\mathrm dV_{bias}} \ln(I) = \frac{n}{V_{bias} - V_{BD}}$ diverges, where $n$ is the model constant which determine the shape of the reverse $IV$. 
Clearly, this divergence is not observed in the experimental data, being a non-physical state. Therefore, from this method one can extract $V_{BD}^{1d}$, a quantity proportional to $V_{BD}$, as the voltage at which the ``relative logarithmic derivative" has a local maximum, for example by fitting the region around $V_{BD}$ with a peaked and skewed function. In our case, we chose a Landau function (see Fig.~\ref{Fig:FirstDerivative}). 

\begin{figure}[tb]
\begin{center}\includegraphics[%
  width=8.2cm,
  keepaspectratio]{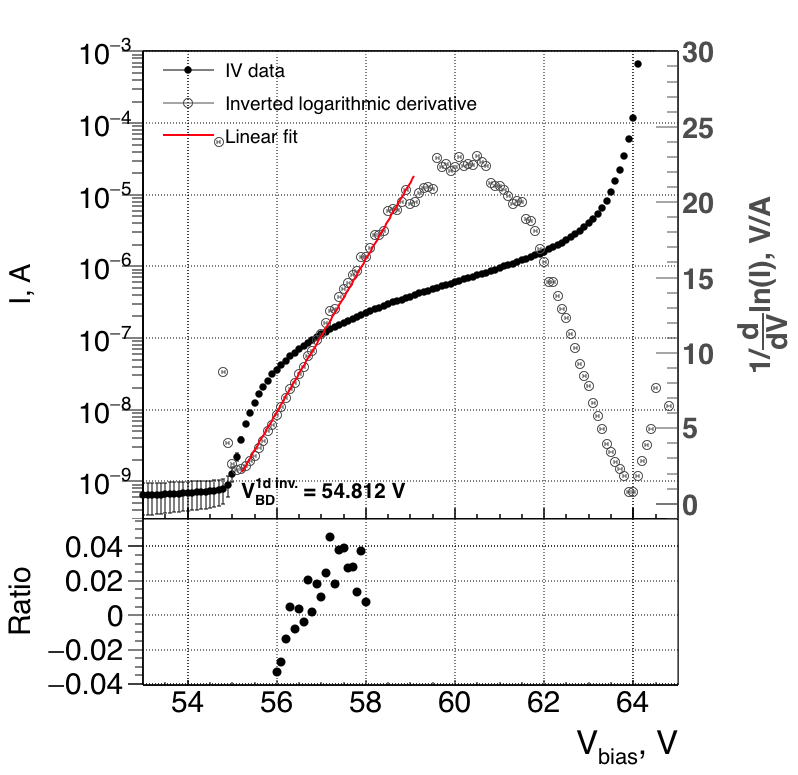}\end{center}
\caption{The reverse IV characteristic data points and its inverted logarithmic derivative. A value of $V_{BD}^{1d inv.} = 54.812$~V is determined as intercept of the x-axis of the right scale and the fitted inverted logarithmic derivative (red line). Also, the $Ratio = \left( 1/ \frac{\mathrm d}{\mathrm dV} \ln(I) - I_{fit} \right) \div 1/ \frac{\mathrm d}{\mathrm dV}\ln(I) $ is shown at the bottom.}
\label{Fig:FirstDerivativeIntverted}
\end{figure}

The ``inverse logarithmic derivative" $1/\frac{\mathrm d}{\mathrm dV_{bias}} \ln(I)$ increases linearly with $V_{bias}$ above the $V_{BD}$ (See Fig.~\ref{Fig:FirstDerivativeIntverted}). Assuming that this behaviour does not change near the breakdown region, the breakdown voltage $V_{BD}^{1d inv.}$ can be extracted as the voltage at which the ``inverse logarithmic derivative'' is equal to zero, i.e. the intersection with the x-axis of the fitted line above $V_{BD}$.

\begin{figure}[tb]
\begin{center}\includegraphics[%
  width=8.2cm,
  keepaspectratio]{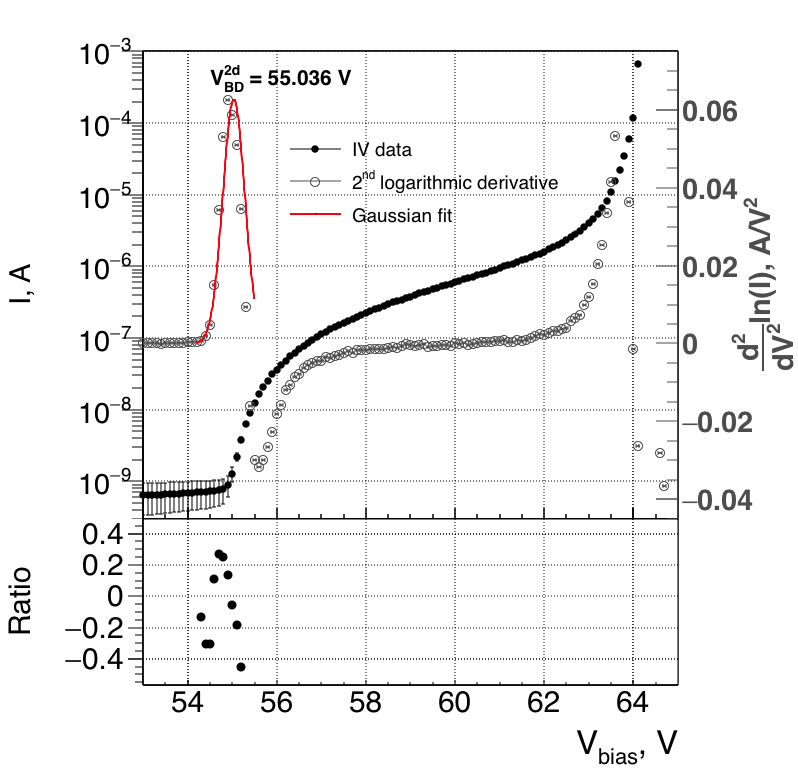}\end{center}
\caption{The reverse IV characteristic data points and its second logarithmic derivative curve. The $V_{BD}^{2d} = 55.036 \ V$ is extracted from the fit of the peak. Also, the $Ratio = \left( \frac{\mathrm d^2}{\mathrm dV^2} \ln(I) - I_{fit} \right) \div \frac{\mathrm d^2}{\mathrm dV^2} \ln(I) $ is shown at the bottom of the figure.}
\label{Fig:SecondDerivative}
\end{figure}

The ``second logarithmic derivative" method~\citep{2ndDerivativeMethod,2rdDerivativeMethod2TCad} is commonly used for the $V_{BD}$ determination.
Here, $V_{BD}^{2d}$ is calculated as the voltage corresponding to the maximum of the second derivative, as shown in Fig.~\ref{Fig:SecondDerivative}. However, we observe that the Gaussian fit does not describe the data well. Therefore, this method determines a systematic error in the absolute value of $V_{BD}$. 

\begin{figure}
\begin{center}\includegraphics[%
  width=8.2cm,
  keepaspectratio]{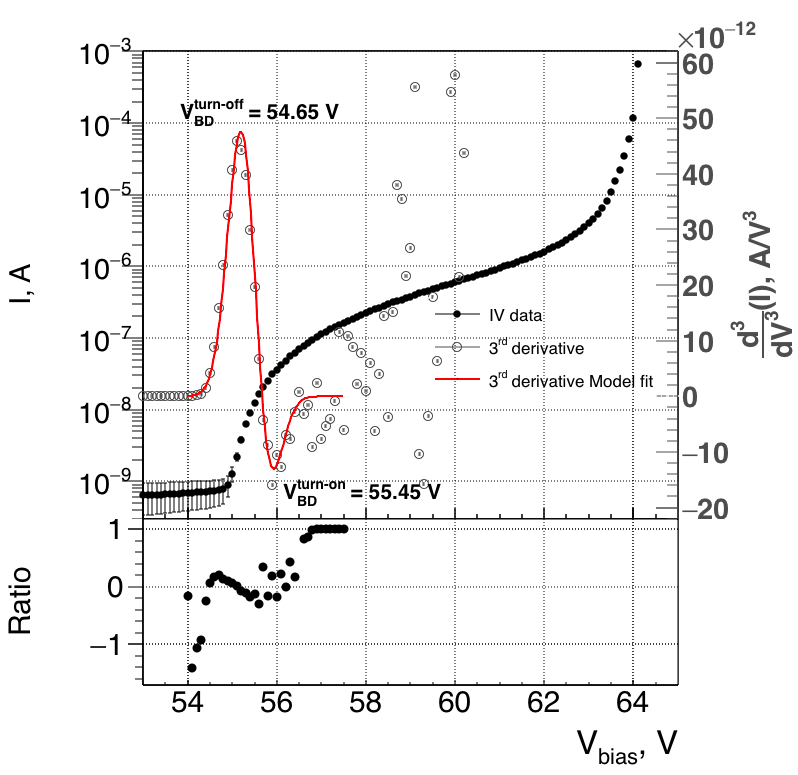}\end{center}
\caption{The reverse IV characteristic data points and its third derivative. The values of 54.65 V and 55.45 V is found for ``turn-off'' and ``turn-on'' $V_{BD}$ from the fit of the curve. Also, the $Ratio = \left( \frac{\mathrm d^3}{\mathrm dV^3}(I) - I_{fit} \right) \div \frac{\mathrm d^3}{\mathrm dV^3}(I) $ is shown at the bottom of the figure.}
\label{Fig:ThirdDerivative}
\end{figure}

The ``third derivative" method~\citep{TurnOnTurnOffVbd,3rdDerivativeMethod} assumes two separate breakdown voltages: the ``turn-on" $V^{3d~turn-on}_{BD}$ and ``turn-off" $V^{3d~turn-off}_{BD}$ voltages. $V^{3d ~turn-on}_{BD}$ defines the regime in which a $\mu$cell initiates an avalanche and the current is related to the avalanche triggering probability $P_{G}$. $V^{3d ~turn-off}_{BD}$ is the voltage at which the quenching of the avalanche starts  and the current is related to charge production. Following the prescription in Ref.~\cite{3rdDerivativeMethod}, we find 54.65 V and 55.45 V for $V^{3d ~turn-off}_{BD}$ and $V^{3d ~turn-on}_{BD}$, respectively (see Fig.~\ref{Fig:ThirdDerivative}).

\begin{figure}
\begin{center}\includegraphics[%
  width=8.2cm,
  keepaspectratio]{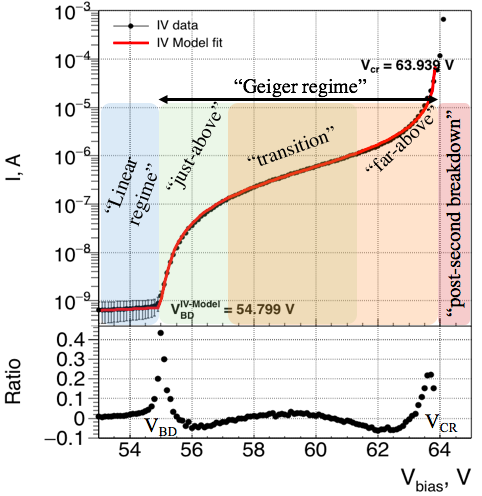}\end{center}
\caption{The reverse IV characteristic data points and its fit done with the ``IV Model'', from which a $V_{BD}^{IV-Model}$ of $54.621 \ V$ is extracted. The main regions (``pre-breakdown", `just-above", ``transition", ``far-above" and ``post-second breakdown") are highlighted with different colours. Also, the $Ratio = \frac{I_{data} - I_{fit}}{I_{data}}$ is shown at the bottom of the figure.}
\label{Fig:IVModelFited}
\end{figure}

To overcome the limitation of all the methods shown so far, a model of the reverse IV curve has been proposed~\citep{IVModeleMethod1,IVModeleMethod2}. 
According to this ``IV Model'', different \sipm{} working regimes can be identified in the IV curve, as shown in Fig.~\ref{Fig:IVModelFited}. As  in Fig.~\ref{Fig:IVreverse} the ``Linear" region (1) is below $V_{BD}$, while here the ``Geiger'' region above $V_{BD}$, is split in four different regions: the  ``just-above'' (breakdown), ``transition'', ``far-above'' and ``post-second breakdown'' zones. 

This  model can describe the IV over the full working range of \sipm{} and therefore it can be used not only to determine breakdown voltage $V_{BD}^{IV-Model}$, but also to determine other \sipm{} parameters such as working range or Geiger probability $P_{G}$ when the IV is measured under light illumination. 
Here we use the procedure described in Ref.~\citep{IVModeleMethod1}, and we obtain a $V_{BD}^{IV-Model} = 54.799$~V.

The systematic uncertainty on all these measurements is given by the:
\begin{itemize}
    \item voltage source accuracy, which has been determined as suggested by the producer~\citep{Keithley2400} as:
        \begin{equation}
            \label{Eq:VbdSys}
            \sigma_{sys.} = V_{BD} \cdot 0.02 \% + 24 \rm mV \sim 35 \rm mV
        \end{equation}.
    \item the model assumptions made to approximate the IV curve with a simple equation:
    \begin{itemize}
        \item as already noted, divergence to infinity of the reverse IV for the ``relative logarithmic derivative" and for the  ``inverse logarithmic derivative" cannot physically observed; 
        \item ``second logarithmic derivative" and ``third derivative" methods do not fit perfectly experimental data;
        \item the "IV Model" does not describe the experimental data near $V_{BD}^{IV-Model}$ and $V_{CR}$. As a matter of fact, a \sipm{} biased below $V_{BD}$ works like an avalanche photodiode and this regime is not included in the IV Model (for more details see Ref.~\citep{IVModeleMethod2}). Additionally, following Ref.~\cite{CHMILL201770}, the $V_{BD}$ value is subject to statistical fluctuations due to  Geiger avalanche statistical fluctuations. On the other hand, the difference near $V_{CR}$ is related to the voltage drop on $R_{q}$. 
    \end{itemize}{}
\end{itemize}{}

\begin{figure}
\begin{center}\includegraphics[ width=8.2cm, keepaspectratio]{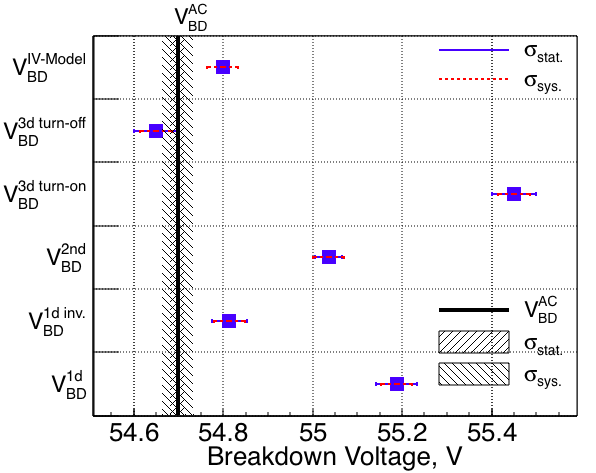}
\end{center}
\caption{$V_{BD}$ with statistics ($\sigma_{stat.}$) and systematic ($\sigma_{sys.}$) errors of the Hamamatsu S10943-2832(X) \sipm{} obtained form static and dynamic measurements.}
\label{Fig:VbdComparison}
\end{figure}

In Fig.~\ref{Fig:VbdComparison}, the  values of $V_{BD}$ obtained using the described methods are compared. They 
are spread over a range of less than 1~V.
In general, the ``inverse logarithmic derivative'' or ``second logarithmic derivative'' methods provide the most stable and straightforward results, and provide a reasonable estimate of $V_{BD}$.
Therefore, those methods are used when many \sipm{} devices should be characterized or compared, as for example in quality assurance procedures.
However, for the full characterization of a device, the "IV model'' should be used, as it can provide the most complete description of the reverse IV. This is the optimal method when design and tuning of front-end electronics is needed or to compare performance of different devices.
Moreover, as will be shown in Sec.~\ref{sec:RelativePDE}, the ``IV Model'' method can also provide the relative photodetection efficiency ($PDE$) of a device.

Fig.~\ref{Fig:VbdComparison} also shows the value of the breakdown voltage as determined with a measurement done with light, which will be described later in Sec.~\ref{sec:AC}.
%\begin{table*}[hbt]
%  \centering
%    \begin{tabular}{c | c | c | c | c | c | c | c}
%    			 & AC  &  \multicolumn{6}{c }{ DC }   \\
%    			  \hhline{~-------}
%              & \multirow{3}{*}{Gain vs. $V_{bias}$}      &   \multicolumn{5}{c |}{ derivative }                             & \multirow{3}{*}{IV Model} \\
%     \hhline{~~-----~}
%               &     &  \multirow{2}{*}{$\frac{d}{dV} ln(I)$}         & \multirow{2}{*}{$1/ \frac{d}{dV} ln(I)$}        & \multirow{2}{*}{$\frac{d^{2}}{dV^{2}}ln(I) (I)$}             &  \multicolumn{2}{c |}{ $\frac{d^{3}}{dV^{3}} (I)$ }             & \\
%    \hhline{~~~~~--~}
%                    &					&				& 				 &						& turn-on & turn-off & \\
%    \hline
%    \hline
%  $V_{BD} (V)$ & 54.699  & 55.188  &  54.814  & 54.960   & 55.45 & 54.65 & 54.799  \\
%  \hline
%   $\sigma_{stat.} (mV)$   & $\pm$ 17 & $\pm$ 46  &  $\pm$ 40 &  $\pm$ 30  & 50 & 50 &  $\pm$ 1 \\
%   \hline
%  $\sigma_{sys.} (mV)$  & \multicolumn{7}{c}{ $\pm$ 25 }   \\
%     \hline
%     \hline
%  $ V_{BD}^{AC} - V_{BD},\ (mV) $ &   & -489 &  -115 & -261 & -751 & 49 & -100\\
%  \end{tabular}
%    \caption{The breakdown voltage obtained from the $Gain$ vs. $V_{bias}$ dynamic (AC) measurement is presented as well as results from static (DC) measurements of the reverse IV curve methods with statistics ($\sigma_{stat.}$) and systematic ($\sigma_{sys.}$) errors for the Hamamatsu S10943-2832(X) \sipm{}. The difference between AC and each DC measurements is also provided.} \label{tab:Comparison}
%\end{table*}

\section{Dynamic characterisation}
\label{sec:AC}

For the dynamic measurements presented here (also refereed further as AC measurements), instead of the standard pre-amplification topology used in the real camera~\citep{SST1Melectronics}(see Fig.~\ref{Fig:PreAmpTopology}), each \sipm{} channel is connected to an operational amplifier OPA846 and readout independently.
The \sipm{} device is illuminated with low intensity light of different wavelengths (e.g. 405  nm, 420 nm, 470 nm, 505 nm, 530 nm and 572 nm) produced by pulsed LEDs.
For each operating voltage of the LED providing a certain light level,
10'000 waveforms are acquired on an oscilloscope and sampled at 500 MHz. Each one is 10~$\rm\mu$s long.  
The signal used to pulse LEDs is produced by a pulse generator and it is also used to trigger waveform acquisition. 

The readout window is adjusted in such a way to have the trigger signal in the middle of the waveform. i.e. at 5~$\rm\mu$s from the window start, in order to have 
\begin{itemize}
\item{} a ``\textit{Dark}'' interval from 0 to 5~$\rm\mu$s, when the device is operated in dark conditions. Only uncorrelated $DCR$ enhanced by correlated noise, i.e. cross-talk (prompt and delayed) and afterpulses, are present (see Sec.~\ref{sec:Noise} for more details);
\item{} ``\textit{LED}'' interval, from 5 to 10~$\rm\mu$s, when the device is illuminated by LED light pulses. In this case, both signals pulses due to the light and uncorrelated \sipm{} noise pulses are present. 
Both types of pulses are further affected by \sipm{} correlated noise (i.e. prompt and delayed cross-talk and afterpulses).
\end{itemize}
%A typical waveform is shown in Fig.~\ref{Fig:WaveformTypical}, where
%Both $Dark$ and $LED$ intervals are marked and \sipm{} pulses from different sources are identified (thermal carriers responsible of $DCR$, LED light, afterpulses, optical prompt and delayed cross-talk).

%\begin{figure*}[bt]
%\centering
%\includegraphics[ width=0.8\textwidth]{WaveformExample.png}
%\caption{Typical waveform of a \sipm{} with signals coming from different sources: thermal carriers, light from LED, after-pulses and optical cross-talk (prompt and delayed).}
%\label{Fig:WaveformTypical}
%\end{figure*}
\textit{Dark} intervals  are used to calculate the \sipm{} $Gain$, the breakdown voltage $V_{BD}^{AC}$, the dark count rate ($DCR$) and the optical cross-talk probability $P_{XT}$, while \textit{LED} intervals are used to calculate the \sipm{} photon detection efficiency $PDE$. To measure the afterpulses probability $P_{AP}$, an additional data run was performed (See sec. \ref{sec:Pap}).

The data acquisition system used for these measurements, consists of a transimpedance amplifier based on OPA846, an oscilloscope Lecroy 620Zi for the waveform acquisition (a bandwidth of 20MHz is used to reduce the influence of the electronic noise) and a  Keithley 6487 to provide bias voltage to the  \sipm{}. For each LEDs of different wavelengths, the over-voltage $\Delta V$ = $V_{bias} \ - V_{BD}^{AC}$ is varied in the range  \mbox{1 V~$< \Delta V < 8$~V}, to cover the full working range of the device (see Sec.~\ref{sec:ReverseIV}).

\subsection{Automatic data analysis procedure}
\label{sec:Analysis}

\begin{figure}[H]
\begin{center}\includegraphics[%
  width=0.8\linewidth,
  keepaspectratio]{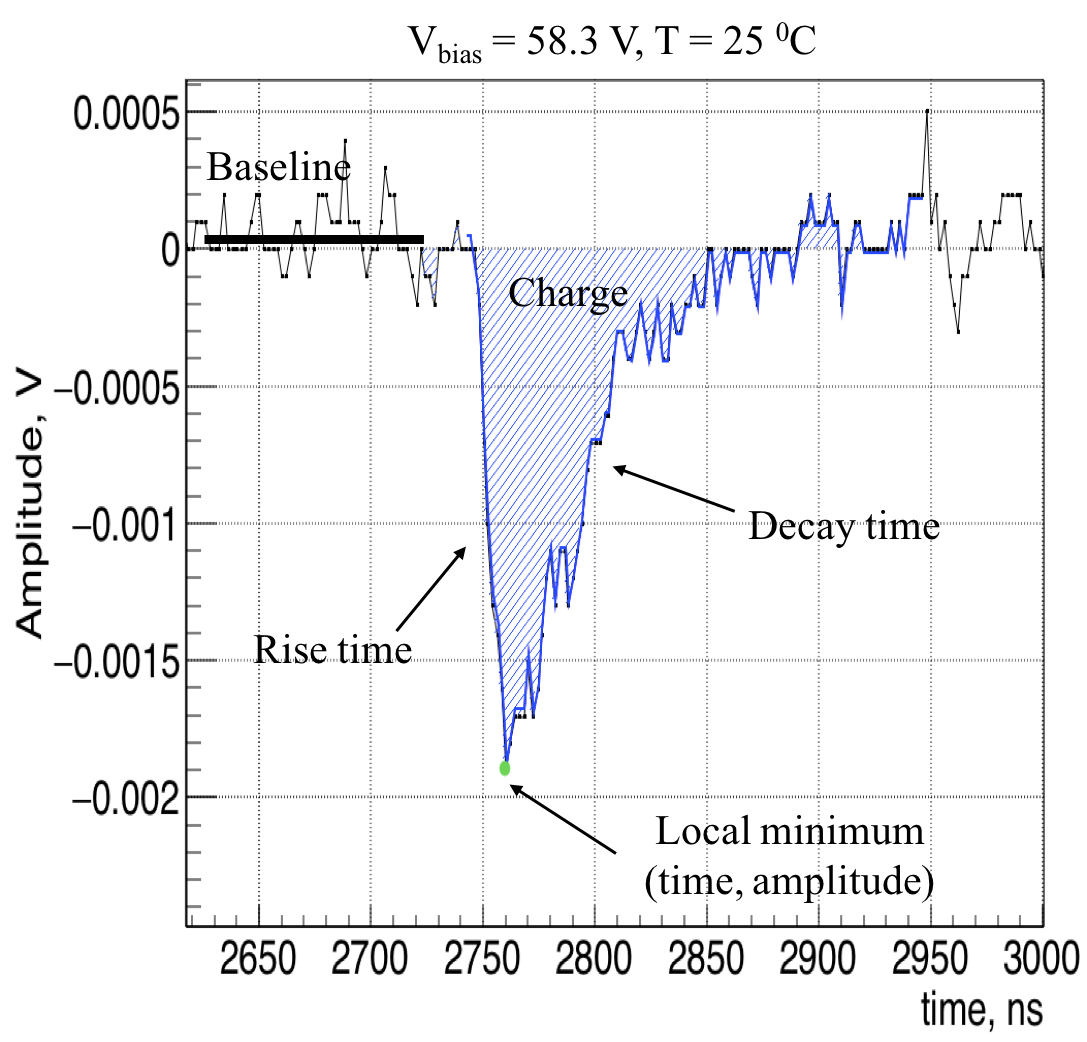}\end{center}
\caption{Typical  \sipm{} pulse for a single photon equivalent (1 p.e.) on top of which its main characteristics are indicated. }
\label{Fig:SignalShape}
\end{figure}

The acquired experimental data are analyzed with an automatic procedure developed in the ROOT Data Analysis Framework~\footnote{\url{https://root.cern.ch}}.
The waveforms acquired with the oscilloscope are used to create ntuples storing \sipm{} pulse templates.
The steps of the analysis to determine the main features of pulses are the following:
\begin{itemize}
  \setlength{\topsep}{0pt}
  \setlength{\itemsep}{0pt}
\item the construction of a template of a typical \sipm{} pulse shape (see  Fig.~\ref{Fig:SignalShape}) in a given working condition;
\item a pulse finding procedure to identify \sipm{} pulses (i.e. a single pulse or a train of pulses~\footnote{By single pulses here it is intended a \sipm{} signal separated by neighbouring pulses by a time interval longer than its recovery time, while  train of pulses is a sequence of two or more signals within a time interval shorter than the \sipm{} recovery time.}) and their relative time spacing;
\item a template subtraction to reconstruct only the \sipm{} pulses in a train of pulses.
\end{itemize}

The \sipm{} pulse characteristics, such as the baseline, time position and amplitude, rise time and decay time, charge $Q$, $t_{before}$\footnote{Time difference between the analyzed pulse and the previous one.} and $t_{after}$\footnote{Time difference between the analyzed pulse and the following one.}, are determined for different values of $V_{bias}$. % and of the temperature T.
More details on the developed analysis procedure can be found in the Ref.\citep{NagaiAnalysis}.

%%%%%%%%%%%%%%%%%%%%%%%%%%%%%%%%%%%%%%%%%%%%%%%%%%%
\subsection{\sipm{} Gain}
\label{sec:Gain}
\begin{figure}[ht]
\begin{center}\includegraphics[%
  width=8.2cm,
  keepaspectratio]{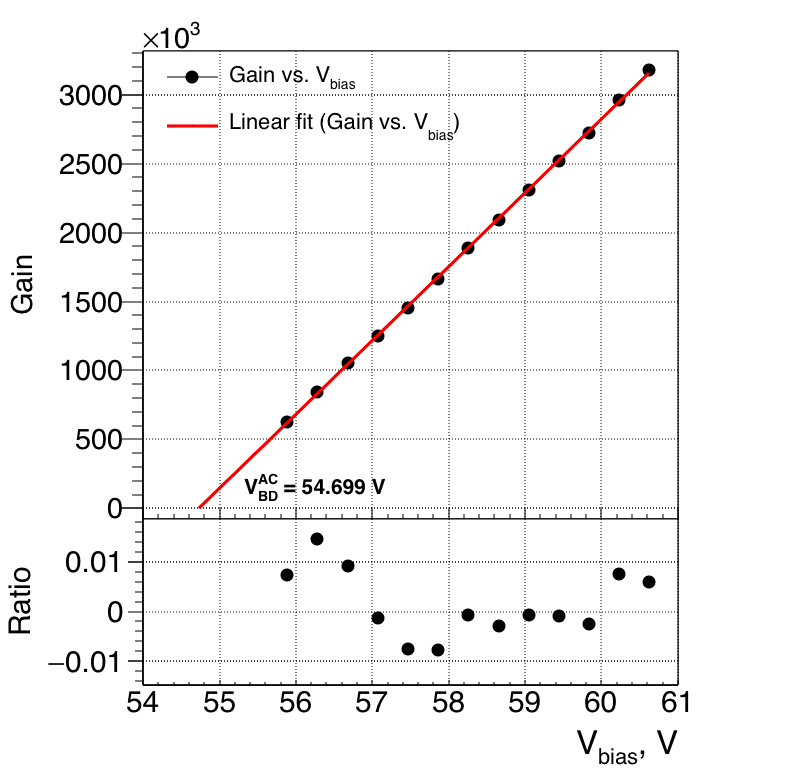}\end{center}
\caption{SiPM gain vs. $V_{bias}$ for the Hamamatsu S10943-2832(X). The $V_{BD}$ of 54.699 V is found at the intersection of the linear fit with the x-axis. Also the ratio, defined as the difference between the experimental data and the fit function values divided by the experimental data, is shown.}
\label{Fig::GanVsVbias}
\end{figure}

%%%%%%%%%%%%%%%%%%%%%%%%%%%%%%%%%%%%%%%%%%%%%%%%%%%
The \sipm{} gain $G$ is defined as the number of charges created by one avalanche in one $\mu cell$ and it can be expressed as:
\begin{equation}
 \label{Eq:Gain}
 G = \frac{Q}{e} = \frac{ \left( C_{\mu cell} + C_{q} \right) \cdot \left( V_{bias} - V_{BD}^{AC}  \right) }{ e },
\end{equation}
where $Q$ is the avalanche charge, $C_{\mu cell}$ and $C_{q}$ are the $\mu cell$ and parasitic capacitance, respectively, and $V_{BD}^{AC}$ is the breakdown voltage (more details are given in Sec.~\ref{sec:Vbd}).
The \sipm{} gain can be calculated from the time integration of the signals of a device:
\begin{eqnarray}
 \label{Eq:GainCalc}
 G = \frac{Q}{e} =
 % TM I think it is enough & \dfrac{1}{G_{Amp} \cdot e}\cdot &\displaystyle \int I(t)dt \\
  \dfrac{1}{G_{Amp} \cdot e} \cdot& \displaystyle\frac{1}{R}\displaystyle \int \left( V(t) - BL \right)dt ,
\end{eqnarray}

where
%$I(t)$ is the current,
$G_{Amp}$ is the amplifier gain, $R$ is the amplifier input impedance ($R = 50 ~\Omega$), $V(t)$ is the pulse evolution over time and $BL$ is the baseline. The gain of the \href{http://www.ti.com/lit/ds/symlink/opa846.pdf}{OPA846} amplifier has strong frequency dependence. Therefore, it will be different for different SiPMs. However, in particularly for our \sipm{} device the $G_{Amp}$ of 5.86 $\pm$ 0.04 was found. At a given $V_{bias}$ and temperature the \sipm{} gain has Gaussian shape as shown in \citep{ECKERT2010217}. Therefore, gain errors were calculated as the errors of the mean.
%and $\int V(t)dt$ is the area under the \sipm{} signal.
As can be seen in Fig.~\ref{Fig::GanVsVbias}, the gain increases linearly with $V_{bias}$ as expected from Eq.~\ref{Eq:Gain}. 

%%%%%%%%%%%%%%%%%%%%%%%%%%%%%%%%%%%%%%%%%%%%%%%%%%%%%%%%%%%
\subsection{Breakdown Voltage
\label{sec:Vbd}}
%%%%%%%%%%%%%%%%%%%%%%%%%%%%%%%%%%%%%%%%%%%%%%%%%%%%%%%%%%%
%The breakdown voltage of a \sipm{} device represents the voltage above which the electrical field inside the depleted region of a $\mu$cell is high enough that any free carrier (created by an absorbed photon or by a thermally generated carrier) can trigger an avalanche.
%%TM Therefore, the $V_{BD}$ determines the left range of the device working region as Geiger Mode Avalanche photo-diode ($GM-APD$).

\begin{figure}[ht]
\begin{center}\includegraphics[%
  width=8.2cm,
  keepaspectratio]{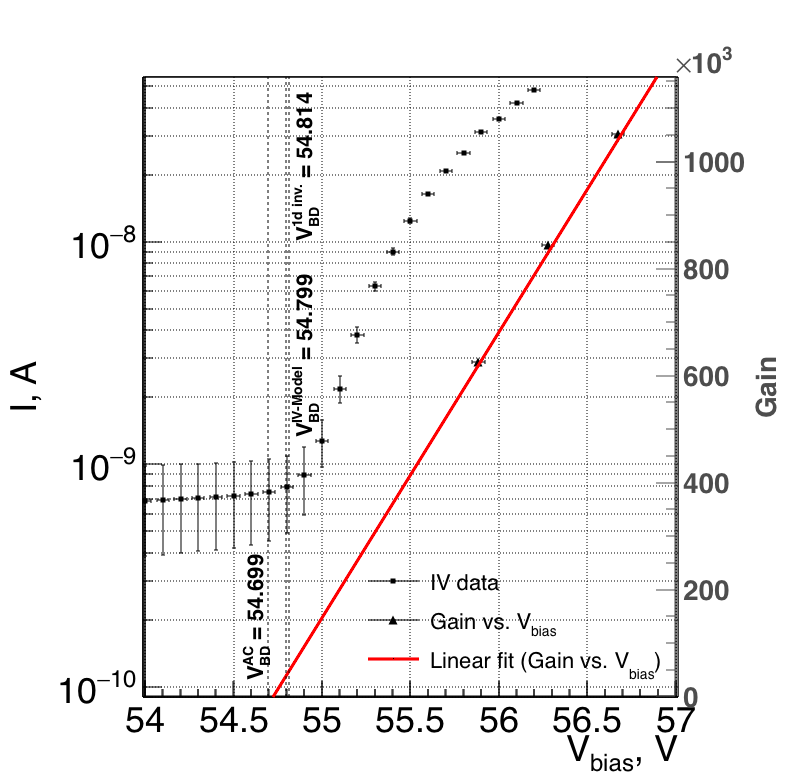}\end{center}
\caption{The zoom near $V_{BD}$ region of the reverse IV curve with superimposed the linear fit of Gain vs. $V_{bias}$ data from Fig.~\ref{Fig::GanVsVbias}. Vertical lines indicate various breakdown voltages: $V_{BD}^{AC}$, $V_{BD}^{IV-Model}$ and $V_{BD}^{1d inv.}$}
\label{Fig::GanVsVbiasWithIV}
\end{figure}

From the curve of the gain as a function of $V_{bias}$  (Fig.~\ref{Fig::GanVsVbias}), the breakdown voltage $V_{BD}^{AC}$ can be determined as the value where $G = 0$ (see Eq.~\ref{Eq:Gain}), i.e.  extrapolating the linear fit to zero.
The obtained value is $V_{BD}^{AC}$ = 54.699 $\pm$ 0.017 (stat.) $\pm$ \mbox{0.035 (sys.) V}. The comparison between this value and those obtained from reverse IV curve static methods (see Sec.~\ref{sec:ReverseIV}) is shown in  Fig.~\ref{Fig:VbdComparison}. 
We can observe significant differences between the dynamic measurement and the static ones, except for the $V_{BD}^{3d turn-off}$ obtained with the $3^{rd}$ derivative method. The $V_{BD}^{AC}$ and $V_{BD}^{turn-off}$ are equal within the uncertainties. However, for all other breakdown voltages, the $V_{BD}^{AC}$ value is a few hundreds of $mV$ smaller than $V_{BD}$ from the IV methods. This discrepancy reflects the described limitation of some of the static methods. 
As it can be seen in Fig.~\ref{Fig::GanVsVbiasWithIV},
the $IV$ static measurement is sensitive to the onset of the avalanche phenomenon and it determines the breakdown voltage, as defined by the fundamental papers of McIntyre (named as ``turn-on" voltage)~\citep{McIntyre}. 
The $Gain$ linearity dynamic method determines the voltage across the diode when the avalanche is quenched (named as ``turn-off" voltage). 
The ``turn-off" (i.e. $V_{BD}^{AC}$ and $V_{BD}^{turn-off}$) is naturally lower than the ``turn-on" (i.e. $V_{BD}^{1d}$, $V_{BD}^{1d inv.}$, $V_{BD}^{2nd}$, $V_{BD}^{3d turn-on}$ and $V_{BD}^{IV-Model}$), as shown in the Fig.~\ref{Fig:VbdComparison}). 

%Nonetheless, a study that could not be verified in this work, compares these two breakdown voltage measurements as function of the cell size shows a different behaviour ~\citep{CHMILL201756}. 
%The breakdown voltage determined with the static methods show to be independent of the cell size within experimental errors, while the one using the gain method shows a significant decrease with decreasing pixel size.

%%%%%%%%%%%%%%%%%%%%%%%%%%%%%%%%%%%%%%%%%%%%%%%%%%%%%%%%%%%
\subsection{\sipm{} micro-cell capacitance and depletion depth}
\label{sec:Cucell}
%%%%%%%%%%%%%%%%%%%%%%%%%%%%%%%%%%%%%%%%%%%%%%%%%%%%%%%%%%%

Combining Eq.~\ref{Eq:Gain} and Eq.~\ref{Eq:GainCalc}, one can extract the device capacitance, which is the sum of the $\mu$cell $C_{\mu cell}$ capacitance and the parasitic $C_{q}$ one. $C_{\mu cell}$ is related to $\mu$cell geometry, through the parallel plane capacitance equation:
%, as show in Eq.~\ref{Eq:CucellCalc}
%\begin{equation}
% \label{Eq:CucellCalc}
% C_{\mu cell} + C_{q} = \frac{G \cdot e}{\Delta V} =  \frac{q}{\Delta V \cdot e} \cdot \frac{1}{G_{Amp}} \cdot \frac{1}{R} \displaystyle \int V(t)dt .
%\end{equation}
%In real life, the amplifier gain $G_{Amp}$ depends  with frequency and value of $C_{q}$ is not known, but it is usually of the order of 3 - 10 fF., for future calculations we assume $C_{\mu cell}$ of 85 fF, as provided by Hamamatsu (See Tab.~\ref{tab:HexSiPM}). 

\begin{equation}
 \label{Eq:Cclasic}
 C_{\mu cell} = \epsilon_{0} \cdot \epsilon_{Si} \times \frac{A}{d},
\end{equation}
where $\epsilon_{0}$ = $8.854 \times 10^{-14}$ F/cm is the vacuum permittivity, $\epsilon_{Si}$ = 11.9 is the silicon dielectric constant, $A = (50 \times 50) \times 0.615$~$\rm\mu$m$^2$ is the active area reduced by the geometrical fill factor of 0.615 and $d$ is the depletion thickness of the $\mu$cell. From this formula, the resulting depletion thickness is $d  = 1.9$~$\mu$m. This relationship between $C_{\mu cell}$ and the depletion thickness was studied with Silvaco TCAD\footnote{\url{https://www.silvaco.com/products/tcad.html}} simulation of the capacitance-voltage characteristic of a diode structure similar to the \sipm{} micro-cell (i.e. p+/n/n-epi/n-substrate). We found agreement between simulated and calculated depletion thicknesses within 0.1 $\mu$m, corresponding to 5.2 $\%$ relative error.

\subsection{\sipm{} noise and $DCR$}
\label{sec:Noise}

\sipm{} noise is a limiting factor for low-light level applications (from one to few photons) and
various mechanisms contribute to it. Two main categories of noise can be identified: the $DCR$ or primary uncorrelated noise, which is independent from light conditions, and the secondary or correlated noise.

%The $DCR$ represents the number of output pulses per second produced when the device is in the dark, determined by few physical phenomenon described below. 
At room temperatures, the $DCR$ is dominated by thermal generation of carriers. 
When a \sipm{} is operated at high overvoltage $\Delta V$ and the electric field across the junction increases, the carriers can tunnel from the valence band to the conduction band through trap or defect states. In this case, the rate of thermally generated carriers is amplified by the trap-assisted tunnelling mechanism.
In addition, the generation rate can be enhanced by the reduction of activation energy due to the Poole-Frenkel effect~\cite{PooleFrankel1938}.
As the electric field increases, the tunnelling of electrons directly from the valence band into the conduction band increases. 
Therefore, at a given temperature, the $DCR$ is determined by the rate of thermally generated carriers $N_{car}$ and the probability that carriers trigger an avalanche (i.e. Geiger probability $P_{G}$). Consequently, a simple empirical formula for the $DCR$ can be approximated as:
\begin{equation}
 \label{Eq:DCRFit}
 DCR = N_{car} \cdot P_{G}^{DCR}\cdot e^{b \cdot V_{bias}} \,
\end{equation}
where $b$ is a free parameter describing the increase of $DCR$ with $V_{bias}$ due to electrical field effects, $P_{G}^{DCR}$ is the average Geiger probability for dark pulses. Following the Refs. \citep{IVModeleMethod1} \citep{OTTE2017106} the Geiger probability can be well expressed as:

\begin{equation}
\label{Eq:Pgeiger}
P_{G} = 1 - e^{-P_{G_{Slope}} \cdot \Delta V}
\end{equation}

where $P_{G_{Slope}}$ is the \sipm{} structural parameter, which determines the rate of increase of $P_{G}$ with $\Delta V$. $P_{G_{Slope}}$ depends on whether an electron or a hole initiates an avalanche (i.e. it detected light of some wavelength) and to some extent on the temperature \citep{COLLAZUOL2011389} (See Sec.~\ref{sec:Pgeiger}). Therefore, $P_{G}^{DCR}$ can be approximated as:

\begin{equation}
\label{Eq:PgeigerDCR}
P_{G}^{DCR} = 1 - e^{-P^{DCR}_{G_{Slope}} \cdot \Delta V}
\end{equation}
where $P^{DCR}_{G_{Slope}}$ is the average $P_{G_{Slope}}$ of $DCR$ pulses.

Secondary, or correlated, noise is due to the optical cross-talk and the afterpulsing induced by a primary avalanche previously generated by a noise source or by detected light photons. 
During the primary avalanche multiplication process, photons can be emitted due to hot carrier luminescence phenomena~\cite{OpticalXtalk}. These photons may lead to:
\begin{itemize}
\item \textit{Prompt optical cross-talk}, due to photons starting  secondary avalanches in one or more neighbour $\mu$cells. 
Therefore, the prompt cross-talk probability $P_{XT}$ can be expressed as:
\begin{equation}
 \label{Eq:PxtFit}
 P_{XT} = G \cdot P_{h \nu}  \cdot P_{G}^{{XT}} 
 %= G \cdot P_{h \nu}  \cdot 1 - e^{-P^{XT}_{G_{Slope}} \cdot \Delta V}
\end{equation}
where $P_{h \nu}$ is the probability that a photons is emitted, reach the high field region of another $\mu$cell and create electron-hole pair, $G$ is the \sipm{} gain, i.e.  the number of charges created during primary avalanche multiplication (see Eq.~\ref{Eq:Gain}) and $P_{G}^{{XT}}$ is the average Geiger probability for cross-talk pulses. $P_{G}^{{XT}}$ can be approximated as:

\begin{equation}
 \label{Eq:PgeigerXT}
 P_{G}^{{XT}} = 1 - e^{-P^{XT}_{G_{Slope}} \cdot \Delta V}
\end{equation}

where $P^{XT}_{G_{Slope}}$ is the average $P_{G_{Slope}}$ of cross-talk pulses.

%is the rate of increase of $P_{G}^{XT}$ with $\Delta V$.

\item \textit{Delayed optical cross-talk}, due to photons, absorbed in the non-depleted regions of the device (i.e. substrate), producing charge carriers that can drift through the depleted region and trigger secondary avalanches ~\citep{PIEMONTE20192, ACERBI201916}. The carrier diffusion time determines the delay time.
\end{itemize}
Afterpulsing occurs when, during the primary avalanche multiplication process,  carriers are captured by trap levels in the $\mu$cell junction depletion layer and are released after some time, triggering a secondary avalanche discharge correlated to the primary one. 
Therefore, the afterpulse probability $P_{AP}$ can be approximated as:
\begin{equation}
 \label{Eq:PapFit}
 P_{AP} = G \cdot P_{trap}  \cdot P_{G}^{P_{AP}} ,
\end{equation}
where $P_{trap}$ is the probability that a carrier will be trapped and released after and $P_{G}^{P_{AP}}$ is the average Geiger probability for afterpulses.

Since, the afterpulsing occurs in the same $\mu$cell as primary avalanche, its amplitude $A_{AP}$ strongly depends on the recovery state of the $\mu$cell, and can be expressed as:

\begin{equation}
 \label{Eq:AfterpulseAmplitude}
 A_{AP}  =  A_{1p.e.} - A_{1p.e.} \cdot \exp \left[  - \frac{ t } { \tau_{rec.} }  \right],
\end{equation}
where $A_{1pe}$ is the single photoelectron (p.e.) amplitude and $\tau_{rec.} = R_{q} \cdot C_{\mu cell}$ is the recovery time constant.

%\subsubsection{Dark count rate, $DCR$}
%\label{sec:DCR}

As mentioned before, the device is considered as operated in dark conditions, in the time  window of about 5 $\mu$s preceding the LED trigger.
This time interval is used to calculate the dark count rate $DCR$.

\begin{figure}[ht]
\begin{center}\includegraphics[%
  width=0.8\linewidth,keepaspectratio]{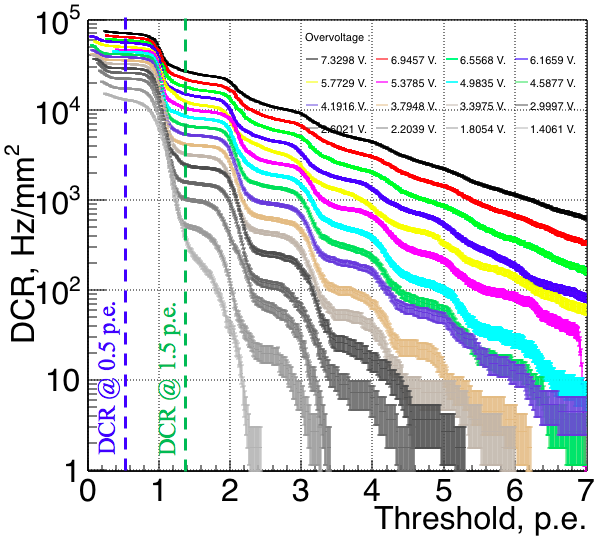}\end{center}
\caption{$DCR$ vs. threshold for different values of overvoltage $\Delta V$ for the Hamamatsu S10943-2832(X) \sipm{}. The blue and green vertical lines represent the DCR at 0.5 p.e. and 1.5 p.e. thresholds, respectively.}
\label{Fig:DCRvsThreshold.png}
\end{figure}

\begin{figure}[ht]
\begin{center}\includegraphics[%
  width=8.2cm,
  keepaspectratio]{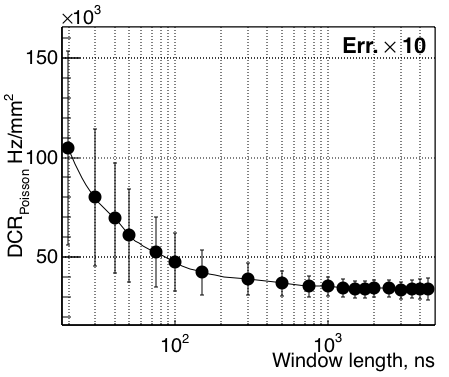}\end{center}
\caption{DCR calculated from Poisson statistics vs analysis window length.}
\label{Fig::DCRPoisVsLength}
\end{figure}

The $DCR$ as a function of a discriminating threshold expressed in photoelectrons $p.e.$ at a given $\Delta V$ is calculated by counting the number of \sipm{} pulses with amplitude above the threshold (see Fig.~\ref{Fig:DCRvsThreshold.png}).
This \textit{counting} method is affected by afterpulses. 
To overcome this limitation, the Poisson statistic can be used to calculate pure uncorrelated SiPM noise at 0.5 p.e. threshold as:
\begin{equation}
    DCR_{Poisson} = -\frac{ ln \left( P_{dark}(0) \right) }{L} =
    - \frac{1}{L}ln \left( \frac{N_{dark}(0)}{N_{dark}(total)} \right)
\label{Eq:DCRPoisson}
\end{equation}
where $P_{dark}(0)$ is the Poisson probability not to have any SiPM pulse and then $-ln(P_{dark}(0))$ is the average number of detected SiPM pulses within the time interval $L$. The $P_{dark}(0)$ can be calculated as:

\begin{equation}
    P_{dark}(0) = - \frac{N_{dark}(0)}{N_{dark}(total)},
\label{Eq:DCRPoissonProb}
\end{equation}
\noindent
where $N_{dark}$(total) represents the total number of analyzed waveform and $N_{dark}$(0) is the number of waveforms without any SiPM pulse within given time interval $L$. 
As can be seen in Fig.~\ref{Fig::DCRPoisVsLength}, the $DCR_{Poisson}$ is overestimated for short window lengths ($\leq$ 1 $\mu$s), as it also affected by afterpulses.
So to estimate correctly the DCR, we need to use a window greater than 1~$\mu$s, where the DCR becomes flat within the error bars.
This value clearly depends on the afterpulse probability and their distribution in time for the specific device, but the same method can be used to identify the right window size for any type of device.

\begin{figure}
\begin{center}\includegraphics[%
  width=8.2cm,
  keepaspectratio]{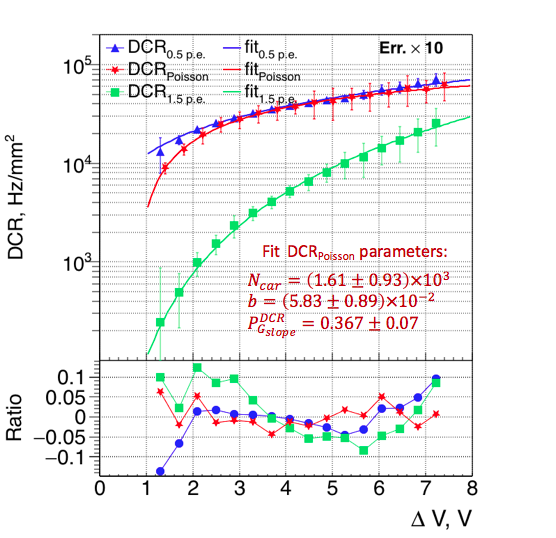}\end{center}
\caption{DCR vs. $\Delta V$ for Hamamatsu S10943-2832(X) for the 0.5 p.e. counting method (blue) as well as from Poisson statistics  method (red) and the 1.5 p.e. threshold (green). Also, the difference between the experimental data and the fit, normalized to experimental data, is presented for the 0.5 (blue), 1.5 (green) p.e thresholds and Poisson statistics (red) are shown. The Fit parameters for the Poisson statistics method are indicated.}
\label{Fig:DCRvsOvervoltage.png}
\end{figure}

Despite the fact that $DCR$ calculated from pulse counting method is slightly overestimated due to afterpulses (see Fig. ~\ref{Fig:DCRvsOvervoltage.png}), it is anyhow interesting to use it to extract other important parameters of the device.

The trend of the $DCR$ measured as function of the overvoltage, at the threshold of 0.5 p.e., can be fitted using Eq.~\ref{Eq:DCRFit}. 
From the fit,  we can extract  $P_{G_{Slope}}^{DCR}= 0.366$ using Eq.~\ref{Eq:PgeigerDCR} (for a physical interpretation of this parameter see Sec.~\ref{sec:Pgeiger}).
The discrepancy between data and fit is larger than the errors. %, as can be seen by the bottom part of Fig.~\ref{Fig:DCRvsOvervoltage.png}.
This can be related to the fact that the fit formula does not include afterpulses and delayed optical cross-talk.
The inclusion of these two effects would make the fit more complex and unstable. This inclusion is not worth given that the errors are quite small (10 ppm at $\Delta = 1.5$~V) and then the discrepancy has negligible impact.

%The $DCR$ at 0.5 p.e. threshold vary from 10 $kHz/mm^2$ at $\Delta V$ = 1.5 V up to 60 $kHz/mm^2$ at \mbox{$\Delta V$ = 7 V}.

\subsubsection{Prompt cross-talk probability
\label{Sec:Pxt}}

The measured $DCR$ at thresholds $\ge$ 1.5. p.e. can be regarded as the results of the optical cross-talk effects related to the $DCR_{0.5 \ p.e.}$ 
and then 
\begin{equation}
 \label{Eq:DCR2petFit}
 DCR_{1.5 \ p.e.} = DCR_{0.5 \ p.e.}
 \times P_{XT},
\end{equation}
which can be used to define how to measure $P_{XT}$: 
\begin{equation}
 \label{Eq:PXT}
 P_{XT} =  \frac{ DCR_{1.5 p.e.} } { DCR_{0.5 p.e.} } .
\end{equation}

However also the pile up effect is present. 
The total rate of pile up pulses within a given time interval $\tau$ can be calculated as the sum of the pile up rate of two, three, four  and more pulses ($R_{total} =  R_{2p} + R_{3p} + R_{4p} + ...$). 
Using a standard approach ~\cite{Grieve:2015aea}, the rate for the estimation of accidental pile up of 2 pulses, with a rate of $DCR_{0.5 p.e.}$ and a coincidence window of $\tau$, is \mbox{$2\cdot \tau \cdot DCR_{0.5 p.e.}^2$}.
Therefore, the total rate, $R_{total}$, for any number pile-up event, can be regarded as a geometrical series of $\tau \cdot DCR_{0.5 p.e.}$:
\begin{eqnarray}
 \label{Eq:PXTPileUp}
 R_{total} &=&  R_{2p} + R_{3p} + R_{4p} + ... \\ \nonumber
 &=&   2 \cdot \tau \cdot DCR_{0.5 p.e.}^{2} +
  2 \cdot \tau^{2} \cdot DCR_{0.5 p.e.}^{3}
  %&+& 2 \cdot \tau^{3} \cdot DCR_{0.5 p.e.}^{4} 
  + ... \\ \nonumber
  &=& \frac{ 2 \cdot \tau \cdot DCR_{0.5 p.e.}^{2} } { 1 - \tau \cdot DCR_{0.5 p.e.}  }\nonumber
\end{eqnarray}

The $P_{XT}$ can be corrected for the pile up effect as:
\begin{equation}
 \label{Eq:PXTCorr}
 P_{XT}^{Corrected}=  \frac{ DCR_{1.5 p.e.}  -  R_{total}} { DCR_{0.5 p.e.} + R_{total}} .
\end{equation}

In our case, the afterpulses can be neglected as they can appear within $\tau = 10\ ns$, and then their contribution to the amplitude is negligible.
As matter of fact, the maximum possible afterpulse amplitude within this time interval was calculated from Eq. ~\ref{Eq:AfterpulseAmplitude} and it is only 0.37 p.e. Therefore, the amplitude of the primary pulse, even including afterpulses, is still below 1.5 p.e. threshold. 

\begin{figure}[ht]
\begin{center}\includegraphics[%
  width=8.2cm,
  keepaspectratio]{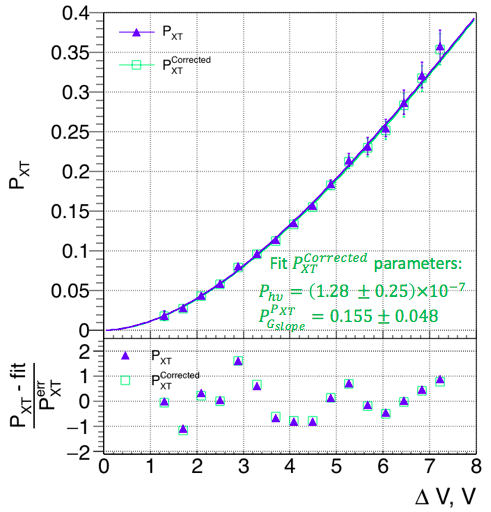}\end{center}
\caption{$P_{XT}$ vs $\Delta V$ of Hamamatsu S10943-2832(X) with ($P^{Corrected}_{XT}$ in green) and without correction for the pile up effect ($P_{XT}$ in blue). In the bottom plot the difference between experimental data and fit normalized to data errors is shown.}
\label{Fig:PXTCorr}
\end{figure}

In Fig.~\ref{Fig:PXTCorr} the prompt optical cross-talk probability, $P_{XT}$, as a function of the over-voltage, $\Delta V$, is shown (blue dots) together with the corrected one (green dots). 
The pile-up correction is below 1\% due to the small value of $\tau$,
which is the minimal separation in time between two pulses needed for the automatic data analysis to recognize them as single pulses inside a train. 
However, the pile-up correction may become important when \sipm{}s with very low $P_{XT}$ are used or when $\tau$ is much longer, as shown in Ref. \citep{NAGAI2018182}.

%The $P^{Corrected}_{XT}$ is approximated by Eq.~\ref{Eq:PxtFit}. 
%We can see that except one point, the difference between fit and data is inside the 5 $\%$ level. 
We can see that except for two points, the difference between fit and data is inside the error bars.
Given $C_{\mu cell} = 85$ fF (See Tab.~\ref{tab:HexSiPM}), two free parameters can be extracted from the fit of $P^{Corrected}_{XT}$: $P_{h \nu}$ and $P_{G_{Slope}}^{P_{XT}}$. 
We find an average probability $P_{h \nu} = 1.28 \times 10^{-7}$ that photon with sufficient energy can be emitted by one carrier crossing the junction during avalanche multiplication and reach the high field region of another $\mu$cell. Taking into account that the average probability of photons with energy higher than 1.14 eV (or $\lambda \leq 1000 \ nm$), emitted by carriers crossing the junction, is $2.9 \times 10^{-5}$ \citep{PxtMeasurements}, we can conclude that around 2\% of emitted photons reach the high field region of neighbouring $\mu$cells. 
The parameter $P_{G_{Slope}}^{P_{XT}}=0.155$ is extracted from the fit. Its physical interpretation will be discussed later in section~\ref{sec:Pgeiger}.

As a further cross-check, we use the value found here for $P_{XT}$ to fit the data in Fig.~\ref{Fig:DCRvsOvervoltage.png} using Eq.~\ref{Eq:DCR2petFit} and Eq.~\ref{Eq:PxtFit}.
The parameters found from the fit of $DCR_{0.5 \ p.e.}$ in Fig.~\ref{Fig:DCRvsOvervoltage.png} are fixed in the fit for $DCR_{1.5 \ p.e.}$. Also in this case, the data are well reproduced by the fitted model, as for $DCR_{0.5~p.e.}$

The $P_{G}^{DCR}$, extracted from $DCR$ data shown in Fig.~\ref{Fig:DCRvsOvervoltage.png}, represents the probability that a free carrier initiates an avalanche (see Eq.~\ref{Eq:DCRFit}), while $P_{G}^{{XT}}$ represents the probability that a photon (emitted by hot carrier luminescence) is absorbed and initiate an avalanche (See Eq.~\ref{Eq:PxtFit}). 
In general, free carriers and luminescence photons are absorbed at different depths of \sipm{} active areas. Therefore, $P_{G}^{DCR} \neq P_{G}^{{XT}}$, even if they have a similar behaviour as a function of $\Delta V$.

%%%%%%%%%%%%%%%%%%%%%%%%%%%%%%%%%%%
\subsubsection{Afterpulse and delayed cross-talk probability}
\label{sec:Pap}

\begin{figure}[ht]
\begin{center}\includegraphics[%
  width=8.2cm,
  keepaspectratio]{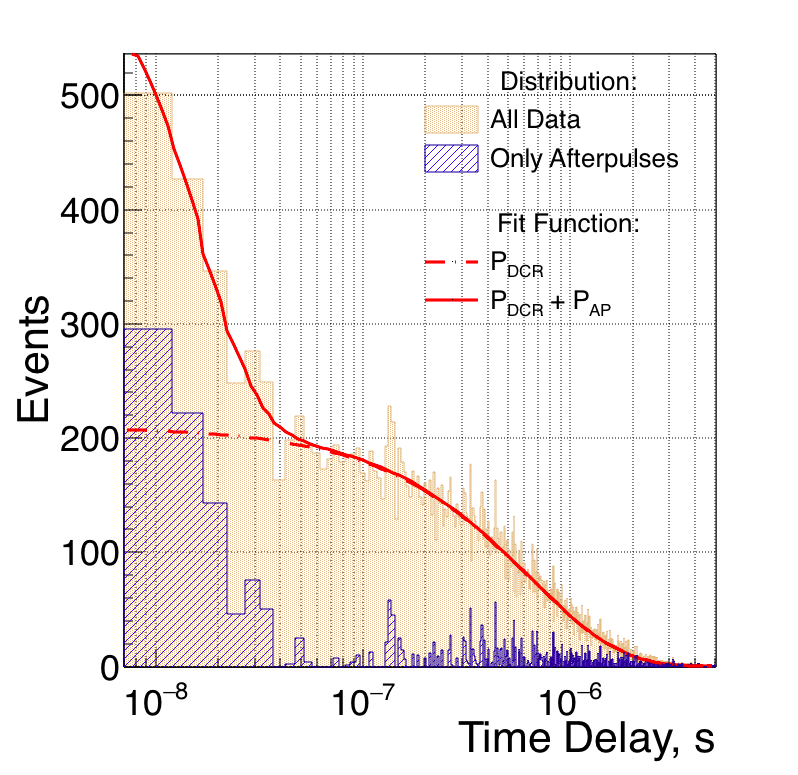}\end{center}
\caption{Distribution of the time difference between primary pulse and first following pulse at $\Delta V$ = 3.5 V. By subtracting the contribution from DCR (dashed line), the distribution for afterpulses only (blue) was obtained.}
\label{Fig:PapTimeDistExample}
\end{figure}

The afterpulse probability is measured by acquiring  20~$\mu$s long waveforms, triggering their acquisition and using a pulse with an amplitude larger than 0.5~p.e..
This pulse, called in the following primary pulse, is adjusted to fall in the center of the waveform (i.e. at  10~$\mu$s). 
To ensure that pulses are either afterpulses related to the primary pulse or randomly generated dark pulses, waveforms without any signal within the 5~$\mu$s preceding the primary pulse are selected and analyzed in the following.  
%Trigger was set in the middle of waveform and triggered pulse will be called in the following primary avalanche. From trigger conditions we can distinguish intervals which contain different data:
%\begin{enumerate}
%    \item from 0 $\mu s$ to 5 $\mu s$ contains $DCR$ pulses enhanced by $P_{XT}$ and $P_{ap}$;
%    \item from 5 $\mu s$ to 10 $\mu s$ doesn`t contain any pulses due to trigger conditions;
%    \item 10 $\mu s$ primary avalanche position;
%    \item from 10 $\mu s$ to 20 $\mu s$ contains afterpulses of primary avalanche enhanced by $P_{XT}$. Also this interval contains $DCR$ enhanced by $P_{XT}$ and $P_{ap}$;
%\end{enumerate}
For all waveforms triggered by a primary signal of 1 p.e. amplitude, the time difference between primary pulse and first following pulse is shown in Fig.~\ref{Fig:PapTimeDistExample}. 
The number of DCR pulses, separated by a given time difference $\Delta t$, can be calculated as:
\begin{equation}
    N_{DCR}(\Delta t) = \frac{ n_{DCR} }{ \tau_{DCR} } \cdot exp \left( \frac{- \Delta t} {\tau_{DCR}} \right),
\label{Eq:NdcrTimeConst}
\end{equation}
where $\tau_{DCR} = 1/DCR$ is the average time difference between two dark pulses and $n_{DCR}$ is the normalization amplitude. Not to include afterpulses, the $DCR$ from the Poisson statistics method was used.
Eq.~\ref{Eq:NdcrTimeConst} is used to fit the data in Fig.~\ref{Fig:PapTimeDistExample}, 
where the contribution due to the afterpulse component is also shown.
This afterpulse component can be approximated as:
\begin{equation}
   N_{AP} = \frac{n_{AP}}{\tau_{AP}} \cdot exp \left( \frac{- \Delta t} {\tau_{AP}} \right)\cdot \left(1 - exp \left( -\frac{\Delta t}{\tau_{rec.}} \right) \right),
\label{Eq:NapTimeConst}
\end{equation}
where $\tau_{AP}$ is the afterpulse time constant and $n_{AP}$ is the normalization amplitude, $1 - exp \left( -\frac{\Delta t}{\tau_{rec.}} \right)$ takes into account decreases of Geiger probability due to micro-cell recovery time. More than one afterpulse time constant (e.g. fast and slow) can presented, as shown in Ref.~\citep{ECKERT2010217} for older \sipm{} devices. For the studied \sipm{}, a single $\tau_{AP}$ was found. This is due to the use of improved materials and wafer process technologies~\citep{HamamatsuBook} reducing drastically afterpulses. Both Eq.~\ref{Eq:NdcrTimeConst} and Eq.~\ref{Eq:NapTimeConst} have similar exponential behavior, even if they are related to different physical 
phenomena: Poisson statistics of \sipm{} uncorrelated noise 
(Eq.~\ref{Eq:NdcrTimeConst}) and \sipm{} trap level lifetime (Eq.~\ref{Eq:NapTimeConst}). 

The data in Fig.~\ref{Fig:PapTimeDistExample} are  approximated as the sum of the 2 components:
\begin{equation}
    N_{total}(\Delta t) = N_{DCR}(\Delta t) + N_{AP}(\Delta t).
\label{Eq:NapDCRtotal}
\end{equation}
This equation neglects the probability $P_{cor}(DCR, AP)$ that after-pulse and $DCR$ pulse may appears in the same micro-cell within the micro-cell recovery time $5 \times \tau_{rec.}$ \citep{Dinu2011}, since it is negligibly small:
\begin{equation}
   P_{cor}(DCR, AP) = \frac{<DCR>}{N_{\mu cell}} \cdot 5 \times \tau_{rec.} \sim 2.5 \times 10^{-5}.
\label{Eq:NapDCRProbabilitySameCell}
\end{equation}
where $<DCR>$ is average $DCR$ over $\Delta V$ at a given $T$.

\begin{figure}[ht]
\begin{center}\includegraphics[%
  width=8.2cm,
  keepaspectratio]{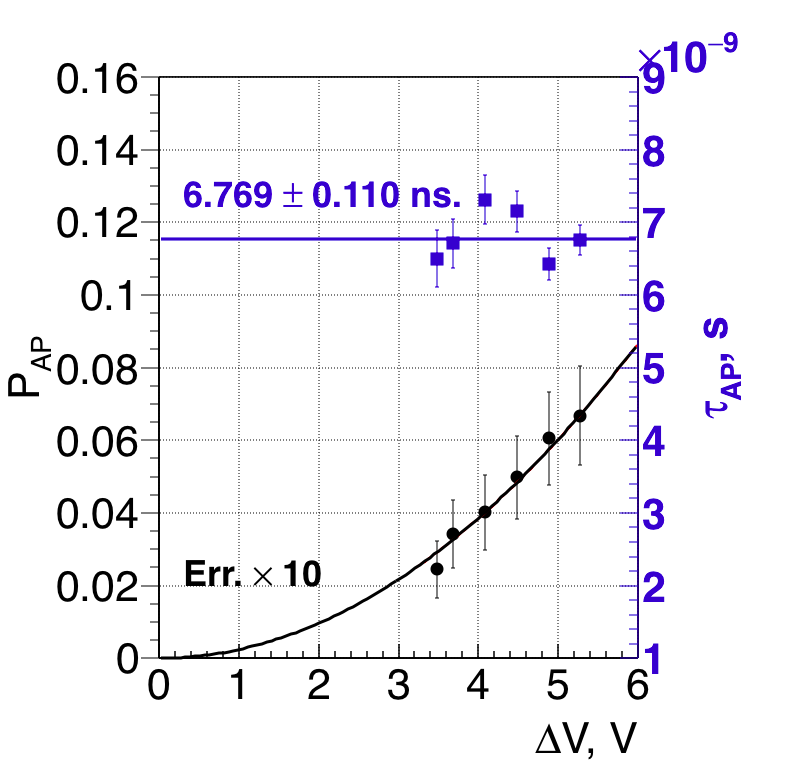}\end{center}
\caption{The $P_{AP}$ and $\tau_{AP}$ as a function of $\Delta V$. The afterpulse average time constant $<\tau_{AP}>$ of 6.769 $\pm$ 0.110 ns is found.}
\label{Fig:PapTimeConstantsVsV}
\end{figure}

Using this approximation to fit the data, the $\tau_{AP}$ can be extracted. It is shown in Fig.~\ref{Fig:PapTimeConstantsVsV} as function of $\Delta V$. Data for $\Delta V \ < \ 3V$ are not presented in Fig.~\ref{Fig:PapTimeConstantsVsV} due to the very low afterpulse probability leading to poor statistics. The $\tau_{AP}$ is a device structure parameter depending on the \sipm{} structure, Si impurities and temperature. Therefore, variations of $\tau_{AP}$ with $\Delta V$ reflect measurements uncertainties. An average value of $<\tau_{AP}>$ of 6.769 $\pm$ 0.110 ns is found.

The number of afterpulses $N_{AP}(\Delta t)$, calculated as the difference between the measured number of events and $N_{DCR}(\Delta t)$, is represented by the blue histogram in Fig.~\ref{Fig:PapTimeDistExample}. Then the afterpulse probability is calculated as:
\begin{equation}
    P_{AP} = \frac{\int_{0}^{5 \times <\tau_{AP}> } N_{AP}(\Delta t) dt}{N_{prim.} },
\end{equation}
where $N_{prim.}$ is the number of primary avalanches. The $P_{AP}$ as a function of $\Delta V$ is presented in Fig.~\ref{Fig:PapTimeConstantsVsV}.
%to the following \sipm{} pulse is stored in and its amplitude are stored in 2d histogram and presented in Fig.~\ref{Fig:Pap2d}. We can observe several populations of pulses: afterpulses, afterpulses enhanced by optical cross-talk, dark counts and dark counts enhanced by optical cross-talk.
%\begin{figure}[H]
%\begin{center}\includegraphics[%
%  width=8.2cm,
%  keepaspectratio]{Afterpulses2d.png}\end{center}
%\caption{2D histogram shows the time difference between primary pulse and following SiMP pulse on the x-axis and the amplitude of the second signal on the y-axis. Several populations can be identified as: afterpulses, $DCR$ both enhanced by optical cross-talk and and are correspondingly labeled.}
%\label{Fig:Pap2d}
%\end{figure}

%Measured probability of the time difference between primary pulse and first following pulse is presented in Fig.~\ref{Fig:PapTimeDistExample}. 

\begin{figure}[ht]
\begin{center}\includegraphics[%
  width=8.2cm,
  keepaspectratio]{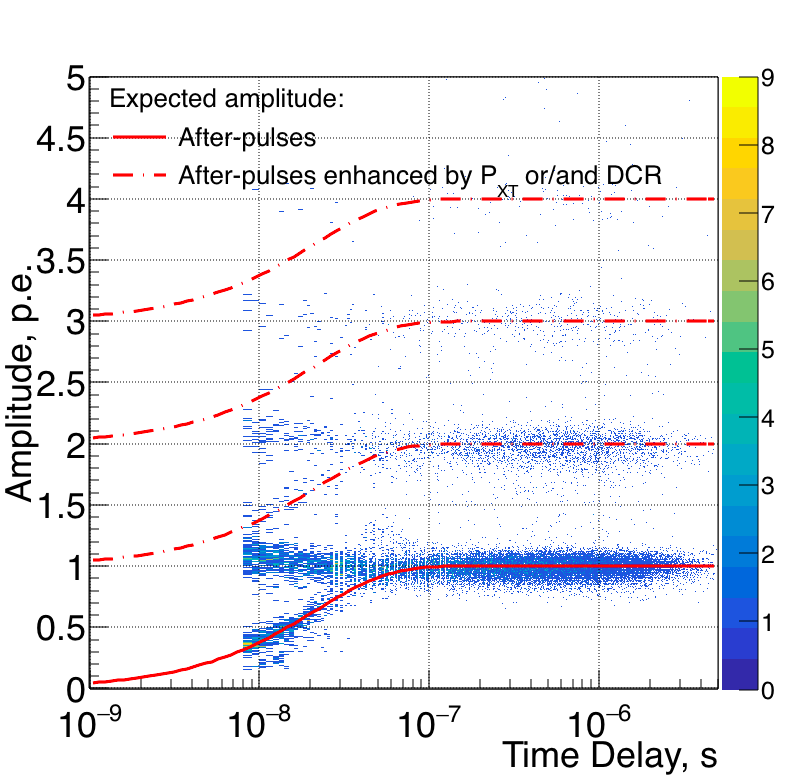}\end{center}
\caption{The 2D histogram shows the time difference between primary pulses and following ones on the x-axis and the amplitude of the second signal on the y-axis. The colors represent the number of events in each bin. The expected afterpulse amplitudes as a function of the delay time is calculated from the $\mu$cell recovery time for pure afterpulses (red solid line) and for enhanced afterpulses by optical cross-talk or dark pulses (dashed red lines).}
\label{Fig:Pap2d}
\end{figure}

Fig.~\ref{Fig:Pap2d} is a two-dimensional histogram of the amplitude in p.e. of the first pulse following a primary pulse of 1 p.e. vs the time difference between the two.
This plot shows the various \sipm{} noise components. The population of dots around amplitude of 1 p.e and time delay larger than 50 ns are typically dark pulses and afterpulses. Nonetheless, for this device only dark counts contribute due to the short afterpulse time constant $\tau_{AP}$. The population with amplitude lower than 1 p.e. and delay smaller than 50 ns are afterpulses produced when the $\mu$cell has not yet recovered. The population at time delay less 50 ns and amplitude 1 p.e. might be mostly delayed optical cross-talk, and some dark pulses or afterpulses related to avalanches happened more than 5 $\mu$s before the primary avalanche. The other populations at larger amplitude than 1 p.e. are of similar nature than what described for 1 p.e. but further enhanced by optical cross-talk. In the plot, the red solid line is calculated from
Eq.~\ref{Eq:AfterpulseAmplitude} and the dashed lines are enhanced by optical cross-talk.

\section{Optical characterisation}
\label{sec:OpticalCharacterisation}

The photon detection efficiency ($PDE$) is one of the most important parameters describing the sensitivity of a \sipm{} as a function of wavelength of the incident  light  $\lambda$ and the applied over-voltage $\Delta V$:
$PDE = QE(\lambda) \times \epsilon \times P_{G}(\Delta V, \lambda)$, where $QE(\lambda)$ is the quantum efficiency, $P_G$ is the Geiger probability, and $\epsilon$ the $\mu$cell fill factor (the percentage of it that is sensitive to light). More details about each $PDE$ component can be found in the Ref.~\cite{ACERBI201916}.
To study the $PDE$, our experimental setup at \href{https://ideasquare.web.cern.ch}{IdeaSquare} at CERN was used (see Fig.~\ref{Fig:SetUp}).
In this Section, the methods used for both absolute (at a given $\lambda$) and relative ($\lambda$-dependent)  $PDE$ measurement are reported and corresponding results discussed at the end of the section.

\subsection{Absolute PDE measurements with pulsed light}
\label{sec:PDEAbsolute}
\begin{figure}
    %\centering
    %\begin{subfigure}[b]{0.4\textwidth}
    %    \centering
    %    \includegraphics[width=\textwidth]{Fig1_set_up.png}
    %    \caption{}
    %    \label{Fig:PhotoSetUp}
    %\end{subfigure}
    %\hfill
    \begin{subfigure}[b]{0.4\textwidth}
        \centering
        \includegraphics[width=\textwidth]{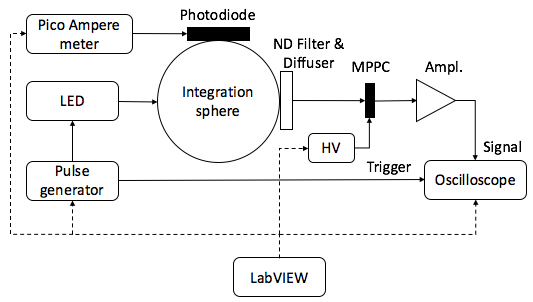}
        \caption{}
        \label{Fig:LEDSetup}
    \end{subfigure}
    \hfill
    \begin{subfigure}[b]{0.4\textwidth}
        \centering
        \includegraphics[width=\textwidth]{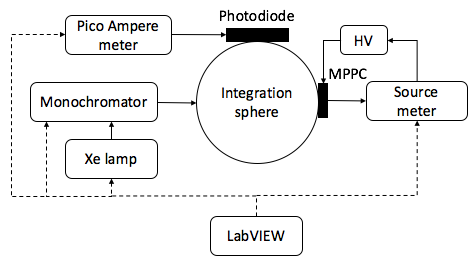}
        \caption{}
        \label{Fig:LampSetup}
    \end{subfigure}
    \caption{Schematic layouts of the developed experimental set-up for: absolute~\ref{Fig:LEDSetup} and relative~\ref{Fig:LampSetup} $PDE$ measurements } \label{Fig:SetUp}
\end{figure}

The schematic layout of the experimental set-up developed for absolute $PDE$ measurements is shown in Fig~\ref{Fig:LEDSetup}. 
The set-up is built around an integration sphere\footnote{Thorlab, Model IS200-4}, used to produce at each output port a diffuse light of similar intensities by multiple scattering reflections on its internal surface. This destroys any spatial information of the incoming light usually produced by a LED but preserves the power at each port.
A calibrated photodiode\footnote{Hamamatsu S1337-1010BQ, s/n 61}, placed on one output port, is used to determine the absolute amount of light scattered in the ports (power density), in order to estimate the number of photons impinging on the \sipm{} under test, sitting on the other port.
The LED bias is provided by a pulse generator, with repetition rate of $f = 500$~Hz, chosen to:
\begin{itemize}
\item have reasonable acquisition time ($\sim 45$~min) for a full scan of the over-voltage in the range of 1 V $ \leq \Delta V \leq$ 8 V with a step of 0.4 V, for each given wavelength; 
\item have a photocurrent level ($I\geq 100$~pA) at least 50 times higher than the \sipm{} dark photocurrent;
\item not saturate the LED, which exhibits a non linear behaviour for $f > 3$~kHz.
\end{itemize}

The dynamic range of the \sipm{}\footnote{The dynamic range is the range where \sipm{} signal charge is linearly proportional to number of photons. As matter of fact, the \sipm{} linearity relies on the fact that each photon hits a different $\mu$cell and the signal is the sum of the charge of the fired $\mu$cells. If the density of photons is too high, the probability that a photon impinges on a $\mu$cell, which has been already fired, and thus is inactive, becomes non negligible. In this case, not all photons contribute to the signal and then the linearity is lost and the device is said to be saturated.} is much lower than the one of a generic photodiode.
To be able to illuminate the \sipm{} with different light intensities, a Neutral Density Filter\footnote{Thorlab, Model NE530B} ($ND \ Filters$) is inserted between the integration sphere output port and the \sipm{}. 
To enable easy and fast replacement, the $ND \ Filter$ is mounted on a motorized wheel. 
To uniformly illuminate the \sipm{} full active area, a $50^\circ \times 50^\circ$  diffuser\footnote{Thorlab, ED1-S50-MD} is mounted after the $ND \ Filter$. 
The surface uniformity is measured using a LED ($\lambda = 405$~nm), and a small photodiode\footnote{Thorlab, Model SM05PD2A} (with 0.8 mm$^2$ active area) mounted on a 2D translation stage\footnote{two Thorlab LTS300 motorized stages connected together by Z-Axis bracket.}. 
The light intensity non-uniformity, which has also to be taken into account for the  $PDE$ calculation, was measured over the active area of the hexagonal \sipm{} and it is $<2\%$. 

%The surface is scanned with a LED of wavelength $\lambda = 405$~nm. For this task the small photo-diode \footnote{Thorlab, Model SM05PD2A} (with 0.8mm$^2$ active area) is fixed on the translation stage. The measurements of the light intensity as a function of $OX$ and $OY$ positions are done over a $30 \times 30$~mm$^2$ area. The normalised light intensity as a function of the $XY$ position is presented in Fig.~\ref{Fig:LightSpotBoth} over the $12 \times 12$ mm$^2$ active area of the \sipm{}. We measure a light intensity non-uniformity of the \sipm{}  of less than $2\%$. To include this non uniformity in the $PDE$ calculation the correction factor for the light non-uniformity $\alpha_{light}$ is calculated as:
% \begin{equation}
%  \label{Eq:LightUniformity}
%  \alpha_{light} =  \frac{ \int_{y_{0}}^{y_{1}} \int_{x_{0}}^{x_{1}} I_{light}(x,y) \ dx dy } { S_{SiPM} } \div \frac{  \int_{ - 5}^{5} \int_{-5}^{5} I_{light}(x,y) \ dx dy }{ S_{PD} } ,
% \end{equation}
%where $I_{light}(x,y)$ is the normalised light intensity at the $x,y$ coordinate (See Fig.~\ref{Fig:LightSpotBoth}), $S_{SiPM}$ and $S_{PD}$ are the active areas of the \sipm{} and of the calibrated photo-diode, respectively, across which the integration is done.

%\begin{figure}
%\begin{center}\includegraphics[%
%  width=8.2cm,
%  keepaspectratio]{LightSpotOne.png}\end{center}
%\caption{Normalised light intensity as a function of $XY$ position.} .
%\label{Fig:LightSpotBoth}
%\end{figure}

The power ratio, $R = P_{PD}/P_{SiPM}$, between the light intensity measured by the calibrated photodiode, $P_{PD}$, and the \sipm{}, $P_{SiPM}$, is measured experimentally as described in Ref.~\citep{BocconeTNS}. Measurements were done for different light wavelengths (i.e. 405, 420, 470, 505, 530, 572 nm). 
%This is done by replacing the \sipm{} by another calibrated photo-diode \footnote{Hamamatsu S1337-1010BQ, s/n 1}, while the diffuser is present not to change the original light spatial position as for the \sipm{}. The measurements done with different LEDs of different wavelengths $\lambda$.

%are presented in Table~\ref{tab:ratio}.
%\begin{table}
%  \centering
%  \begin{tabular}{ c | c }
%    Wavelength (nm) & $R$  \\ \hline
%    405 & 1.26 $\times 10^{-3}$   \\ \hline
%    420 & 1.35 $\times 10^{-3}$ \\ \hline
%    470 & 1.35 $\times 10^{-3}$\\ \hline
%    505 & 1.37 $\times 10^{-3}$ \\ \hline
%    530 &  1.39 $\times 10^{-3}$ \\ \hline
%    572 &  1.38 $\times 10^{-3}$
%  \end{tabular}
%      \caption{Measured power ratios $R$ for different LED wavelengths.}
%\label{tab:ratio}
%\end{table}

\begin{figure}
\begin{center}\includegraphics[%
  width=8.2cm,
  keepaspectratio]{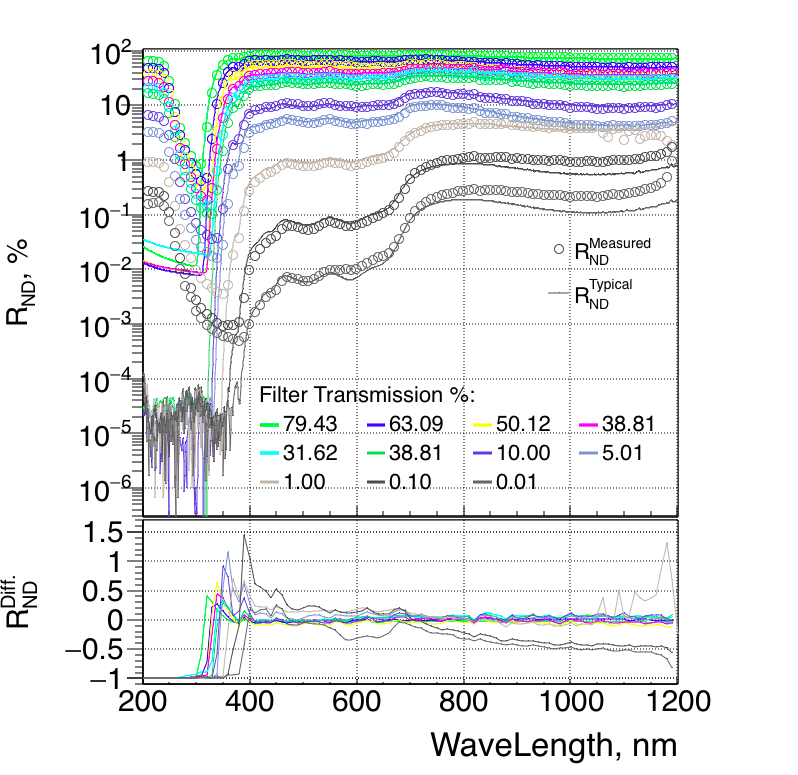}\end{center}
\caption{Measured (open dots) $R^{Measured}_{ND}(n)$ and "typical" one $R^{Typical}_{ND}(n)$ given by producer (lines) transmission of ND Filters as a function of wavelength. In the bottom pane,  it is shown the relative differences $R^{Diff.}_{ND}(n)$ as defined in the text.}
\label{Fig:NDTransmission}
\end{figure}

%In order to correct for this effect in the PDE measurement, 
The transparency ($R_{ND}(n)$) of the $ND \ Filter$ $n$ at a given $\lambda$ is measured as:
\begin{equation}
 \label{Eq:NDFilterTransparency}
 R_{ND}(n) =  
 \frac{I_{ND}(n) } { I_{PD} }  \cdot \frac{1}{R_{Geom.}} ,
\end{equation}

where $n$ is the $ND \ Filter$ number ($n = 1$ is used when there is no filter); $I_{ND}(n)$ and $I_{PD}$ are the photocurrents measured by one photodiode positioned after the $ND \ Filter$ and the reference photodiode positioned at another output of the integration sphere, respectively; $R_{Geom.} = \frac{I_{ND}(n = 1)}{I_{PD}}$ is the power ratio between the light intensity measured by the photodiodes when there is no $ND \ Filter$. In order to measure $R_{ND}$, a Xenon lamp (75 W) was coupled with a monochromator\footnote{Oriel Tunable Light Source System TLC-75X} to select $\lambda$. 
The comparison between the measured values,  $R^{Measured}_{ND}(n)$, and the ''typical" one given by the producer, $R^{Typical}_{ND}(n)$, as a function of $\lambda$ and for different attenuation filters is presented in Fig.~\ref{Fig:NDTransmission} 
together with the relative differences, on bottom of the figure.
%TM trivial definition
%\begin{equation}
% \label{Eq:NDFilterTransparencyDiff}
%R^{Diff.}_{ND}(n) = \frac{ R^{Typical}_{ND} (n) - R^{Measured}_{ND}(n) }{ R^{Measured}_{ND}(n)},
%\end{equation}
%For further measurements the $ND \ Filter$ NE530 was used.

The data acquisition system is similar to the one presented in Sec.~\ref{sec:Analysis}.
During data taking, the photocurrent of the photodiode is read out by the Keithley 6487. Data taking is triggered by a pulse generator and controlled by a Labview program to automate the necessary measurement steps.

The absolute $PDE$ is calculated using the so-called Poisson method \citep{ECKERT2010217,DinuPDE,BocconeTNS, ACERBI201916} from the average number of detected photons, corrected by factor $k^{corr}_{LED}$ to take into account the uncorrelated \sipm{} noise: 

%For every waveform, the local minimum $A_{min}$ of the waveform amplitude (we are working with negative pulses) is calculated within a time window in which the light from the LED is expected to be contained (the $LED \ gate$ of $\simeq 40 ns$). A typical distribution of $A_{min}$ for waveforms acquired at $V_{bias}$ = 59.2 V is presented in Fig.~\ref{Fig:AminDist}. Due to reasons already described in Sec.~\ref{sec:Pap}, the ``0 p.e'' peak is selected to determine the average number of detected photons $k_{LED}$ eliminating any influence from cross-talk (prompt and delayed) and after-pulses:

%A typical distribution of $A_{min}$ for waveforms acquired at $V_{bias}$ = 59.2 V is presented in Figure~\ref{Fig:AminDist}. We can observe various peaks, each of them corresponding to a given number of fired microcells (e.g. detected 1 p.e., 2 p.e. and 3 p.e.). The number of photons from LED is expected to follow the Poisson distribution. However, the observed distribution of p.e. is distorted due to optical cross-talk and after-pulses. Nevertheless, these phenomena do not affect the 0 p.e. peak. Therefore, 0 p.e. peak indicates the number of recorded waveforms $N_{LED}(0)$ with no \sipm{} signal and consequently it can be used to determine the average number of detected photons $k_{LED}$ without any influence from these two effects (e.g. cross-talk and after-pulses):

%\begin{multline}
% \label{Eg:Poisson}
% P(n_{p.e}) = \frac{k_{LED}^{n_{p.e.}}}{ n_{p.e} !} \times e^{-k_{LED}} \underset{n_{p.e. = 0}} \Rightarrow \\
% \Rightarrow k_{LED} = - \ln \left( P_{LED}(0) \right) = - \ln \left( \frac{N_{LED}(0)}{N_{LED}(total)}  \right) ,
%\end{multline}
%where $P(n_{p.e.})$ is the probability to detect a given number of photo electrons $n_{p.e.}$, $N_{LED}(total)$ is the total number of events.
%However, thermal \sipm{} pulses can appear within the $LED \ gate$ used for the $A_{min}$ calculation. Therefore the value of $k_{LED}$ is corrected for dark \sipm{} pulses using an auxiliary $dark \ gate$. with equal length with respect to the $"LED" \ gate$, but at the beginning of a waveform, when the \sipm{} is operated in dark conditions):
%\footnote{formula is not correct..bad cut \& Paste}

\begin{eqnarray}
 k_{LED}^{corr} &=& - \ln \left( P_{LED}(0) \right) + \ln \left( P_{dark}(0) \right)   \nonumber  \\
&=& - \ln \left( \dfrac{N_{LED}(0)}{N_{LED}(total)}  \right) + \ln \left( \dfrac{N_{dark}(0)}{N_{dark}(total)}  \right) ,
 \label{Eq:kLEDCorrected}
\end{eqnarray}
\noindent
where $N_{dark}(0)$ and $N_{dark}(total)$ are the number of waveforms with no \sipm{} signal within the $dark$ regions preceding it and the total number of recorded waveforms, respectively.

%\begin{figure}
%\begin{center}\includegraphics[%
%  width=8.2cm,
%  keepaspectratio]{AminHist.png}\end{center}
%\caption{Local minimum of a waveform amplitude calculated within the $LED \ gate$, the time gate during which the pulses from the LED are expected.}
%\label{Fig:AminDist}
%\end{figure}

\begin{figure}
\begin{center}\includegraphics[%
  width=8.2cm,
  keepaspectratio]{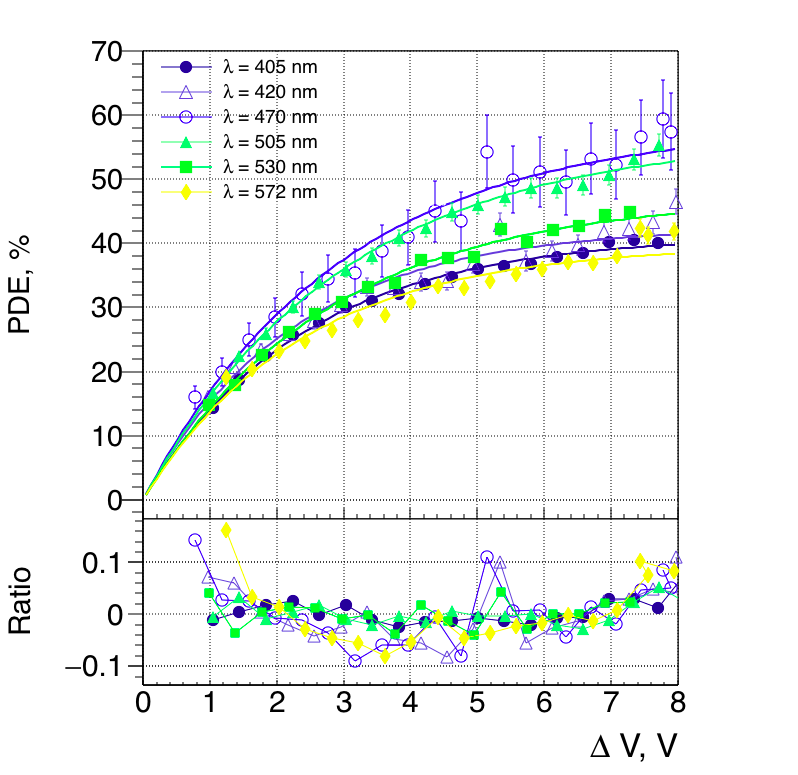}\end{center}
\caption{$PDE$ vs. $\Delta V$ of the Hamamatsu S10943-2832(X) \sipm{}. The results are presented for six different wavelengths: 405 nm, 420 nm, 470 nm, 505 nm, 530 nm and 572 nm. Also, the $Ratio = \left( PDE_{data} - PDE_{fit} \right) \div PDE_{data}$ is shown.}
\label{Fig:PDEvsOvervoltage}
\end{figure}

The $PDE$ can be calculated as:
\begin{equation}
 \label{Eq:PDESimple}
 PDE = \frac{k_{LED}^{corrected}}{N_{ph}} ,
\end{equation}
where $N_{ph}$ is the average number of photons hitting the \sipm{}. The $N_{ph}$ can be estimated from converting the photocurrent from the calibrated photodiode as:
\begin{equation}
 \label{Eq:Nphoton}
 N_{ph} = \frac{I_{PD} \times R \times R_{ND} \times  \alpha_{light} }{f \times QE_{PD}(\lambda) \times e} ,
\end{equation}
where $I_{PD}$ is the photocurrent measured by the calibrated photodiode, $QE_{PD}(\lambda)$ is the photodiode quantum efficiency, $f$ is the pulse repetition frequency (typically $f = 500 \ Hz$) and $e$ is the electron charge.
The PDE as a function of $\Delta V$ for six different wavelengths is shown in Fig.~\ref{Fig:PDEvsOvervoltage}. There are three main  sources of uncertainty for the PDE determination:
\begin{itemize}
\item The precision on  $N_{ph}$, calculated from the photodiode current, used for its calibration curve, and corrected by the power ratio $R = P_{PD}/P_{SiPM}$.
\item The determination of $k_{LED}^{corr}$ based on the separation of the ``0 p.e." and ``1 p.e." peaks, affected by \sipm{} noise (i.e. $DCR$, $P_{XT}$ and $P_{AP}$), which are proportional to the $"LED" \ gate$.
\item The precision of the calibrated quantum efficiency curve of the  photodiode. 
\end{itemize}
Therefore, for precise absolute $PDE$ measurements perfectly calibrated photodiodes, fast LEDs or lasers are strongly preferable. 
In the Fig.~\ref{Fig:PDEvsOvervoltage}, we can observe that the error bars are different for different wavelengths. 
This reflects the variations of the LED light intensity during the measurements, which determine the precision of $k_{LED}^{corr}$ calculation.

The $PDE$ of a \sipm{} can be obtained fitting the data as a function of $\Delta V$ (see Fig.~\ref{Fig:PDEvsOvervoltage}) for each wavelength with the function:
\begin{equation}
 %PDE &=& PDE_{max} \times P_{G} = \nonumber \\
 %&=& PDE_{max} \times \left[ 1 - exp \left( P_{G_{Slope}} \times \Delta V \right) \right],
 PDE = PDE_{max} \times P_{G}
 \label{Eq:PDEFit}
\end{equation}
where $P_{G}$ is the Geiger probability (See. Eq. \ref{Eq:Pgeiger}) and  $PDE_{max}$ is  a free parameter, which
depends on SiPM type, light wavelength and to some extent on temperature ~\cite{COLLAZUOL2011389}.
Such a parameterisation provides a good description of our experimental data as shown in Fig.~\ref{Fig:PDEvsOvervoltage}.

\subsection{Relative {\it PDE} measurement with continuous light}
\label{sec:RelativePDE}

The absolute $PDE$ measurements method requires a pulsed light source, as LEDs or a laser, so it is possible only for a limited number of wavelengths. Therefore, to measure the $PDE$ in a wide wavelength range, from 260~nm up to 1150~nm, a second method, the so called ``Relative PDE'', is used. 
The schematic layout of the experimental set-up developed for the relative $PDE$ measurement is shown in Fig~\ref{Fig:LampSetup}.
%where a 75 W Xenon lamp is used as light source and it is coupled to a mono-chromator is to select the wavelengths in a short range. 
The reverse current-voltage IV characteristics of the \sipm{} device at different wavelengths are performed using a Keithley 2400, while a Keithley 6487 is used to read photocurrent from calibrated photodiode.

%\begin{figure}
%\begin{center}\includegraphics[%
%  width=8.2cm,
%  keepaspectratio]{ContiniousSchematic.png}\end{center}
%\caption{Schematic layout of the relative PDE measurement setup with continuous light.}
%\label{Fig:LampSetup}
%\end{figure}

\begin{figure}
\begin{center}\includegraphics[%
  width=8.3cm,
  keepaspectratio]{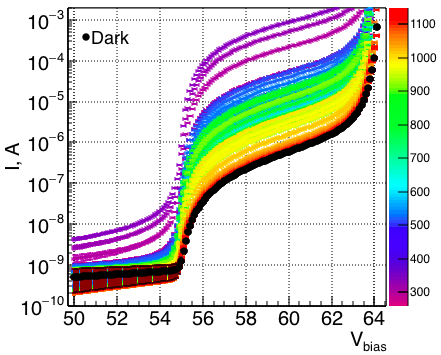}\end{center}
\caption{The reverse IV measurements of the Hamamatsu S10943-2832(X) \sipm{} illuminated by continuous light for various wavelengths from 260 nm up to 1150 nm. 
and in dark (black dots). The colors represent the wavelengths}
\label{Fig:IVvsWavelength}
\end{figure}

The collection of reverse IV curves of the Hamamatsu S10943-2832(X) \sipm{},  for different wavelengths from 260~nm up to 1150~nm, is shown in Fig.~\ref{Fig:IVvsWavelength}. 
The difference between \sipm{} current with light and in dark condition $I_{SiPM}^{light} - I_{SiPM}^{dark}$ at a given $\Delta V$ can be expressed as:

\begin{equation}
 \label{Eq:ISiPM}
I_{SiPM}^{light} - I_{SiPM}^{dark} = PDE(\Delta V, \lambda) \times N_{\gamma} \times e \times G_{SiPM}^{eff.}(\Delta V),
\end{equation}
where $PDE(\Delta V, \lambda)$ is the $PDE$ at a given $\Delta V$ and $\lambda$, $N_{\gamma}$ is the average number of photons sent to the \sipm{} device per given time interval, $G_{SiPM}^{eff.}(\Delta V)$ is the effective \sipm{} gain, namely the \sipm{} gain enhanced by cross-talk and afterpulses effects (for more details see Sec.~\ref{sec:Analysis}). 
The $N_{\gamma}$ is proportional to the photocurrent from the calibrated photodiode $I_{PD}(\lambda)$. Therefore, the relative $PDE$ in Eq.~\ref{Eq:ISiPM} can be rewritten as:

\begin{equation}
 \label{Eq:PDERelative}
PDE(\Delta V, \lambda) = \frac{ I_{SiPM}^{light} - I_{SiPM}^{dark} }{ e \times N_{p.e.} \times G_{SiPM}^{eff.}(\Delta V) } \propto \frac{ I_{SiPM}^{light} - I_{SiPM}^{dark} }{ I_{PD} (\lambda) }
\end{equation}

\begin{figure}
\begin{center}\includegraphics[%
  width=8.2cm,
  keepaspectratio]{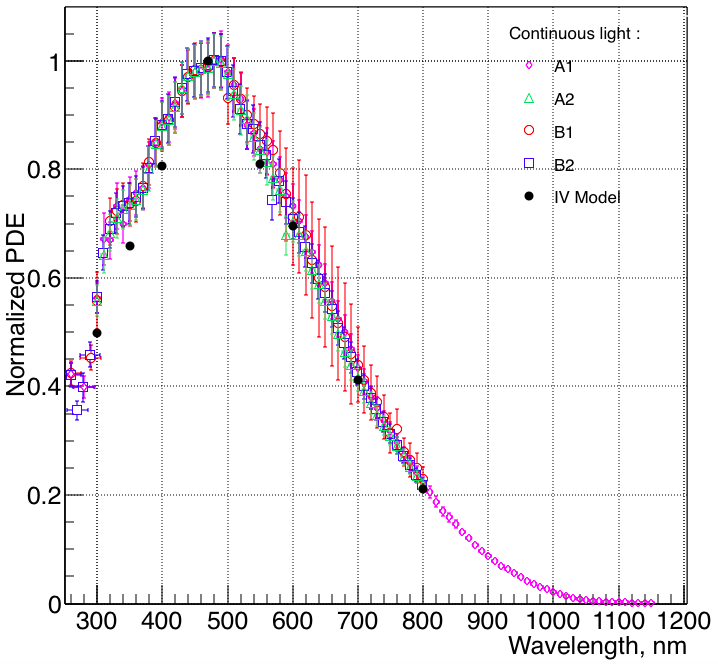}\end{center}
\caption{The relative $PDE$ vs $\lambda$ for the Hamamatsu S10943-2832(X) \sipm{}. The results are presented for all four channels: A1, B1, A2, B2 and also for the $PDE$ calculated from the ``IV Model" for the channel A1 (black points).}
\label{Fig:RelativePDEvsWavelength}
\end{figure}

The relative $PDE$ as a function of $\lambda$  at $\Delta V = 2.8$~V is presented in Fig.~\ref{Fig:RelativePDEvsWavelength}, together with the values  as calculated from the ``IV Model'' (see Sec.~\ref{sec:ReverseIV}) by re-normalising them to the light intensity as estimated with the calibrated photodiode. 

At a given temperature,  the $C_{\mu cell}$ and $\mathrm dN_{car}/\mathrm dt$ of a \sipm{} device  do not depend on light intensity, but only on the \sipm{} internal structure. 
Therefore, a simultaneous fit is done 
%by Eq.~\ref{Eq:FitModel}, 
assuming that $C_{\mu cell}$ and $\frac{\mathrm dN_{car}}{\mathrm dt}$ are the same for all curves. 
To reduce computing time, the fit procedure used only  eight curves corresponding to 300, 350, 400, 470, 550, 600, 700 and 800 nm wavelengths. 
The relative $PDE$ calculated from the ``IV Model'' is in good agreement with the results calculated from Eq.~\ref{Eq:PDERelative}, as shown in Fig.~\ref{Fig:RelativePDEvsWavelength}. 
The main advantage  of the ``IV Model'' for relative $PDE$ calculation is that also the breakdown voltage is extracted from the fit.
As a matter of fact, in  Eq.~\ref{Eq:PDERelative} the currents are measured as function of $V_{bias}$ and then to derive the $PDE$ vs over-voltage,  the $V_{BD}$ has to be known or determined independently.

\begin{figure}
\begin{center}\includegraphics[%
  width=8.2cm,
  keepaspectratio]{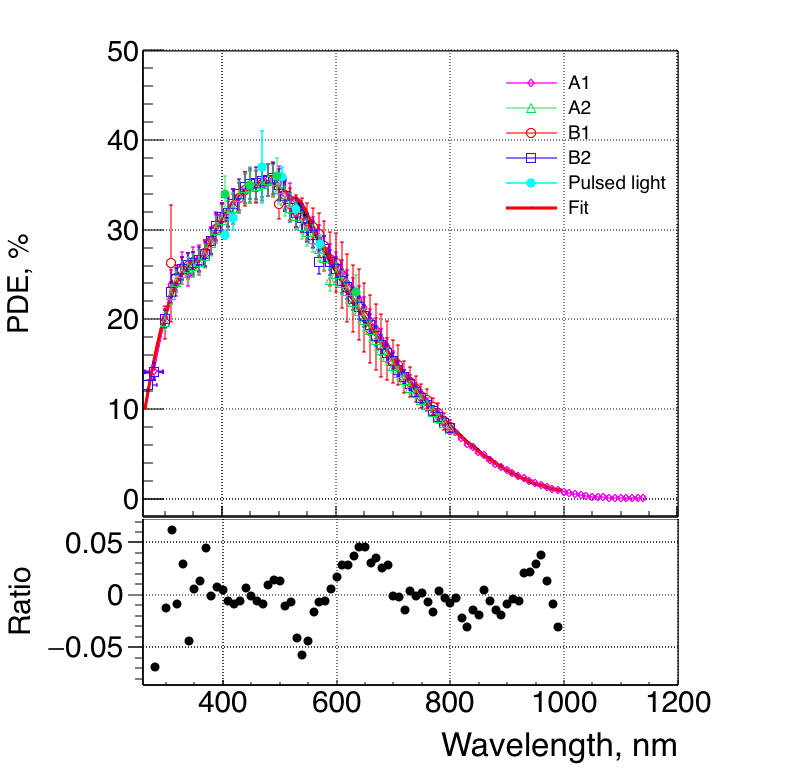}\end{center}
\caption{The PDE vs wavelength for the Hamamatsu S10943-2832(X) \sipm{} from 260 nm up to 1000 nm at $\Delta V = 2.8 V$. Each channel is presented by a different colour and the PDE from pulsed light is presented by cyan dots. Also the fit is shown by the red line and the $Ratio = \left( PDE_{data} - PDE_{fit} \right) \div PDE_{data}$ is shown in the bottom plot.}
\label{Fig:PDEvsWavelength}
\end{figure}

To have an absolute $PDE$ vs $\lambda$, the relative $PDE$ is normalised to the absolute values obtained from Eq.\ref{Eq:PDEFit} at  $\Delta V = 2.8$~V and presented in Fig.~\ref{Fig:PDEvsWavelength}. 
Due to the complicated behaviour of the $PDE$ vs $\lambda$, a sum of three polynomial functions is used to fit the experimental data from 260 up to 1000~nm:

\begin{eqnarray}
PDE(\lambda)& = &\left( a_{1} + b_{1} \cdot \lambda + c_{1} \cdot \lambda ^{2} \right) \cdot  \mathcal{H}_{1}(\lambda) \nonumber\\
&+&\left( a_{2} + b_{2} \cdot \lambda + c_{2} \cdot \lambda ^{2} \right)  \cdot \mathcal{H}_{2}(\lambda) \\
&+&\left( a_{3} + b_{3} \cdot \lambda + c_{3} \cdot \lambda ^{2} + d_{3} \cdot \lambda ^{3}  \right)  \cdot \mathcal{H}_{3}(\lambda),\nonumber
\label{Eq:PDEvsWavelengthFit} 
\end{eqnarray}

% \begin{multline}
%  \label{Eq:PDEvsWavelengthFit}
% PDE(\lambda, \Delta V = Cons.) = \left( a_{1} + b_{1} \cdot \lambda + c_{1} \cdot \lambda ^{2} \right) \cdot \\
%  \cdot \mathcal{H}_{1} (\lambda)
% \begin{cases}
%   1 & 260 \ nm < \lambda < 370 \ nm \\
%   0 & \lambda < 260 \text{ nm or } \lambda > 370 \text{ nm}
% \end{cases}
%  + \\ + \left( a_{2} + b_{2} \cdot \lambda + c_{2} \cdot \lambda ^{2} \right) \cdot \\
%  \cdot \mathcal{H}_{2}(\lambda)
% \begin{cases}
%   1 & 370 \ nm < \lambda < 530 \ nm \\
%   0 & \lambda < 370 \text{ nm or } \lambda > 530 \text{ nm}
% \end{cases}
% + \\ + \left( a_{3} + b_{3} \cdot \lambda + c_{3} \cdot \lambda ^{2} + d_{3} \cdot \lambda ^{3}  \right) \cdot \\
% \cdot \mathcal{H}_{3}(\lambda)
% \begin{cases}
%   1 & 530 \ nm < \lambda < 1000 \ nm \\
%   0 & \lambda < 530 \text{ nm or } \lambda > 1000 \text{ nm}
% \end{cases}
% \end{multline}
\noindent 
where $a_{i}, b_{i}, c_{i}$, with $i = 1,2,3$, and $d_{3}$ are free parameters and $\mathcal{H}_{i}$ are Heaviside step functions in the ranges:\\
$\mathcal{H}_{1}: 260$~nm $< \lambda <$ 370~nm,
$\mathcal{H}_{2}: 370$~nm $< \lambda <$ 530~nm, \\
$\mathcal{H}_{3}: 530$~nm $< \lambda < 1000$~nm.

The result in Fig.~\ref{Fig:PDEvsWavelength} shown a good agreement between experimental data and the fit using Eq.~\ref{Eq:PDEvsWavelengthFit}.

\begin{figure}
\begin{center}\includegraphics[%
  width=8.2cm,
  keepaspectratio]{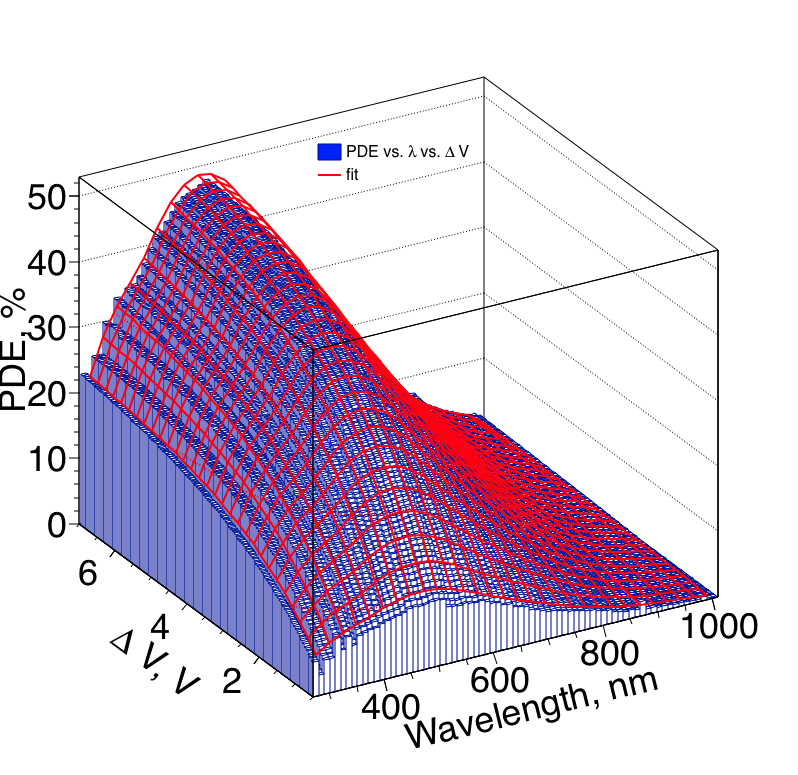}\end{center}
\caption{The PDE vs. $\lambda$ and $\Delta V$ for S10943-2832(X).}
\label{Fig:PDEvsWavelengthvsOverV}
\end{figure}

The $PDE$ as a function of $\Delta V$ and $\lambda$ is particularly useful to predict its variations with experimental conditions, such as temperature or the NSB \citep{VdropHeller}, which affects $\Delta V$ and then the sensor response.
The $PDE$ as a function of $\Delta V$ and $\lambda$, shown in Fig.~\ref{Fig:PDEvsWavelengthvsOverV},  can be obtained by combining the the absolute and relative $PDE$  measurements.

\begin{figure}
\begin{center}\includegraphics[%
  width=8.2cm,
  keepaspectratio]{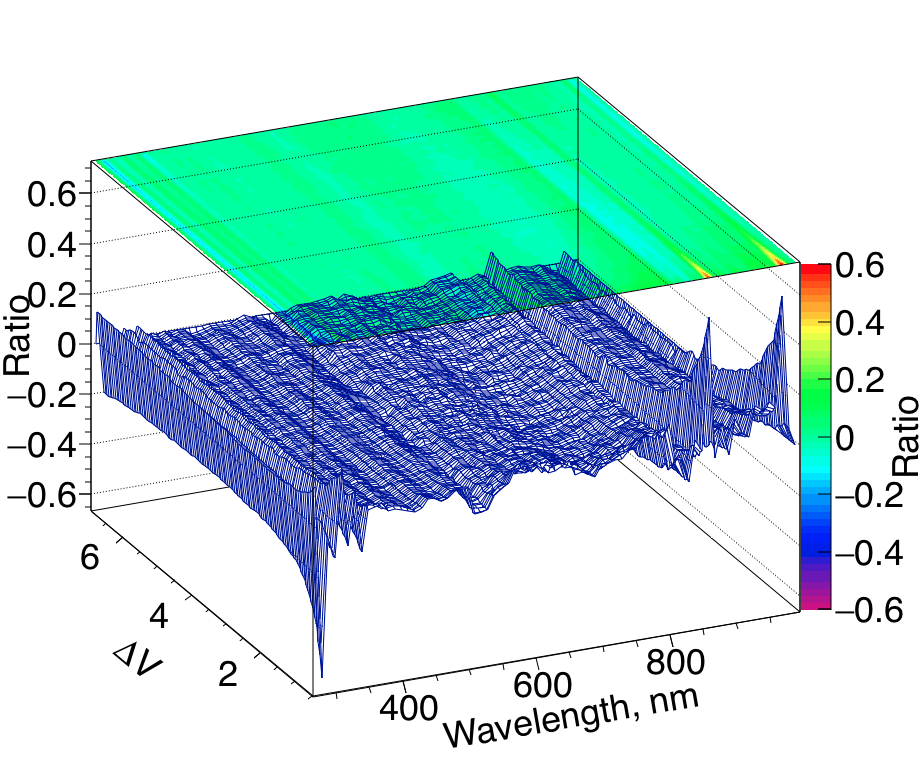}\end{center}
\caption{The difference between the measured PDE and the fit function divided by the measured PDE.}
\label{Fig:PDERatioOverV}
\end{figure}

%In fact, the absolute $PDE$ can be calculated as function of the over-voltage from Eq.\ref{Eq:PDEFit} but for a limited number of wavelengths, while the relative $PDE$, which can also calculated as function of the over-voltage below the ``second breakdown" from IV curves, can also provides the $PDE$ as function of the wavelength, 
%
The analytical expression of the $PDE$ is given in Eq.~\ref{Eq:PDEFit}.
%where $PDE_{max}(\lambda)$ is the free parameter defining the $PDE$ saturation and given by Eq.~\ref{Eq:PDEvsWavelengthFit}, $P_{G}(P_{G_{Slope}}^{DCR}(\lambda, \Delta V) )$ is the Geiger probability (See Eq.~\ref{Eq:FitModePgeiger}). The_{G_{Slope}}^{DCR}$ as a function of $\lambda$ is well represented by a polynomial of $3^d$ order. 
As can be seen in Fig.~\ref{Fig:PDERatioOverV}, data and this representation agree within 3\% on average. 
At low over-voltages ($\Delta V \leq 1.5$~V), there is the largest disagreement between the fit function and the data:
\begin{itemize}
\item  $\lambda \leq 300$~nm, the Xe lamp was operated with a larger slit width of 1.24~mm, to have enough light. As consequence, the wavelength resolution was of  16.1~nm and this resulted in lower precision on $PDE$ (See Fig.~\ref{Fig:PDEvsWavelength}) and then in a worse quality fit;
\item for $\lambda \geq 800$~nm, the photocurrent generated by the \sipm{} is comparable to its dark current (see Fig.~\ref{Fig:IVvsWavelength}) and, therefore, the signal to noise ratio is low.  
\end{itemize}

%\begin{figure}
%\begin{center}\includegraphics[%
%  width=8.2cm,
%  keepaspectratio]{PDESlopeNew.png}\end{center}
%\caption{The $P_{G_{Slope}}^{DCR}$ as a function of $\lambda$. Also, the $P_{G_{Slope}}^{DCR}$ obtained from the $DCR$ and the $P_{XT}$ measurements are presented.}
%\label{Fig:PDESlopevsWavelength}
%\end{figure}

\subsection{Geiger probability
\label{sec:Pgeiger}}

The Geiger probability $P_{G}$, also known as triggering probability, represents the probability that a carrier reaching the high field region will trigger an avalanche. 
%As shown in Eq.~\ref{Eq:PGeiger}, $P_{G}$ can be expressed as function of $P_{G_{Slope}}$.
%This last parameter as function of $\lambda$ was tested to be  well reproduced by a polynomial of $3^d$ order, which was used in the fit of $PDE$ vs. $\lambda$ and $\Delta V$.
%In Fig.~\ref{Fig:PDESlopevsWavelength}, is shown the $3^d$ order function resulting from the parameters extracted form the fit.

\begin{figure}
\begin{center}\includegraphics[%
  width=8.2cm,
  keepaspectratio]{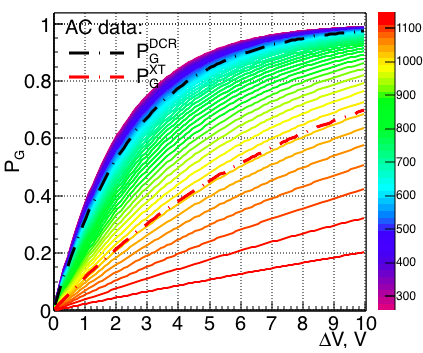}\end{center}
\caption{$P_{G}$ as a function $\Delta V$ for various $\lambda$. The colors represent the wavelengths. Also the average $P_{G}$ for $DCR$ and for $P_{XT}$ are shown.}
\label{Fig:PGeigervsWavelength}
\end{figure}

The $P_{G}$ as a function of $\Delta V$ for different wavelengths $\lambda$ is shown in Fig.~\ref{Fig:PGeigervsWavelength}, where it is evident  how $P_{G}$ increases with increasing $\Delta V$ much rapidly for short wavelengths (blue light). Oldham \citep{Oldham} and McIntyre \citep{McIntyre} relate this behaviour to the properties of light absorption in silicon and to the \sipm{} $\mu$cell structure and ionisation rates of electrons, $\alpha_{e}$. and holes, $\alpha_{h}$. 
In particular, for $p^{+}/n/n-epi/n$-sub-structure
%, as the one   represented schematically in Fig.~\ref{Fig:DiodeStr}, 
$P_{G}$ at short wavelengths $\lambda$ (blue light) is dominated by $\alpha_{e}$ while at long one (red light) it is dominated by $\alpha_{h}$.
Thus, the fast increase of $P_{G}$ with $\Delta V$ at short $\lambda$ is related to the fact that $\alpha_{e} >> \alpha_{h}$ \citep{CrowellSze}.

%Following Oldham \citep{Oldham}, the $P_{G}$ is a combination of electrons $P_{e}$ and holes $P_{h}$ triggering probabilities:
 %\begin{equation}
 %\label{Eq:PGeigerOldham}
%P_{G} = P_{e} + P_{h} - P_{e} \times P_{h}
%\end{equation}
%The $p^{+}/n/n-epi/n-sub$ structure \sipm{} has two boundary conditions:
 %\begin{equation}
 %\label{Eq:PGeigerBoundary1}
%P_{e}(d) = 0
%\end{equation}
 %\begin{equation}
% \label{Eq:PGeigerBoundary2}
%P_{h}(0) = 0
%\end{equation}
%where $d$ is the depletion thickness of micro-cell.

%From these boundary conditions we can assume that $P_{G}$ at short $\lambda$ (blue light) represented only by $P_{e}$, while $P_{G}$ at long $\lambda$ (red light) represented only by $P_{h}$. Following Oldham \citep{Oldham} and McIntyre \citep{McIntyre} $P_{e}$ increases with increasing $\Delta V$ much faster with respect to $P_{h}$ (i.e. $\alpha_{e} >> \alpha_{h}$) as we can observe in Fig.~\ref{Fig:PGeigervsWavelength}.

The average probabilities that thermal pulses (see Sec.~\ref{sec:Noise}) or pulses created by optical cross-talk (see Sec.~\ref{Sec:Pxt}) trigger an avalanche are indicated as $P_{G}^{DCR}$ and $P_{G}^{{XT}}$ in Fig.~\ref{Fig:PGeigervsWavelength}. As can be seen,
$P_{G}^{DCR}$ is equal to $P_{G}$ (black dashed line) at $\lambda$ = 565 nm and  $P_{G}^{{XT}}$ (red dashed line) is equal to $P_{G}$ at $\lambda$ = 1041 nm. %, as can be derived from $P_{G_{Slope}}^{DCR}$in Fig.~\ref{Fig:PDESlopevsWavelength}.
%Knowing that the absorption depth in Si is about  1.7 $\mu m$ for 565 nm and about 440 $\mu m$ for 1041 nm.  
From this and previously discussed behavior between $\lambda$ and $\alpha_{e}$, $\alpha_{h}$ we may conclude that:
\begin{itemize}
%\item the main contribution of $DCR$ is coming from carriers thermally generated in the $n-epi$ layer. 
%In fact,  the measured device has a $p^{+}/n/n-epi/n-sub$ micro-cell structure of 1.9 $\mu m$ depletion thickness (see Sec.~\ref{sec:Cucell}), so the 1.7 $\mu m$ depth should correspond to the $n-epi$ layer (see green line in Fig.~\ref{Fig:DiodeStr});
%\item the main contribution of $P_{XT}$ is coming from photons emitted in hot carrier luminescence phenomena, which penetrate down to the end of the $n-sub$ layer, where they are reflected back into the active area and trigger secondary avalanches (400 $\mu m$ correspond to the tipical \sipm{} thickness as see in Fig.~\ref{Fig:DiodeStr});
\item the main contribution of $DCR$ is triggered by both electrons and holes; 
\item the main contribution of $P_{XT}$ is triggered by holes;
\end{itemize}

\section{Conclusions}

In this work we report about the characterization measurements of the large area hexagonal \sipm{} S10943-2832(X). 
We measure all relevant \sipm{} parameters, detailed in Tab.~\ref{tab:HexSiPMFinal}.
We also show how to build a PDE function of the wavelength and over-voltage. This is of paramount importance to determine the working point of \sipm{}s in real applications where external factors can affect its parameters, such as in the presence of NSB.
Additionally, we compare several methods commonly used for $V_{BD}$ estimate from the reverse current voltage $IV$ measurement.
The functions to fit $PDE$, $DCR$ $P_{XT}$ and $P_{AP}$ are discussed. We also show how, from these fits, 
the triggering probability $P_{G}$ as function of the wavelength can be extracted. From its behaviour we infer that the DCR is triggered by both electrons and holes, while the cross-talk is initiated by avalanches triggered mainly by holes.
%, therefore photons emitted in hot carrier luminescence phenomena are reflected in the $n$-$sub$ layer into the active area.

\begin{table}[hbt]
  \centering
  \renewcommand{\arraystretch}{1.4}
\begin{tabular}{||r|c||}
\hhline{|t:==:t|}
Breakdown voltage $V_{BD}^{AC}$ &  54.699 $\pm$ 0.017 $\pm$ 0.025 V\\
$DCR/mm^{2} \ @ \ 0.5 p.e.$ (@$V_{op}$) 	& 26.50 $\pm$ 0.15 KHz\\
$DCR/mm^{2} \ @ \ 1.5 p.e.$ (@$V_{op}$) 	& 1.735 $\pm$ 0.04 KHz\\
$P_{XT}$ (@$V_{op}$)	&  6.5 \%\\
$P_{AP}$ (@$V_{op}$ within 5~$\mu$s) & $<$ 2  \% \\
PDE (@$V_{op} \ \& $  $\lambda$ = 472~nm ) &	35.5 $\pm$ 3.5 \%\\
Peak sensitivity wavelength & 480~nm \\
Quenching resistor $R_{q}$ & 182.9 $\pm$ 0.3 $\pm$ 31. $k\Omega$\\
\hhline{|t:==:t|}
  \end{tabular}
\caption{S10943-2832(X) \sipm{} main measured characteristics at T = 25 $^{\circ}$C. $V_{op} = V_{BD} + 2.8$ V.}
\label{tab:HexSiPMFinal}
\end{table}
\section*{Acknowledgements}
This project has received funding from the European Union$\textsc{\char13}$s Horizon 2020 research and innovation programme under grant agreement No 713171. Also, we acknowledge the support of the SNF funding agency. This paper has gone through internal review by the CTA Consortium.

\section*{References}
\bibliography{SST-1M}
\end{document}